%% file: ozdes.tex
\documentclass[fleqn,usenatbib]{mnras}

\usepackage{newtxtext,newtxmath,gensymb}

\usepackage[T1]{fontenc}
\usepackage{ae,aecompl}

\AtBeginShipout{%
  \ifnum\value{page}>1 %
    \typeout{* Additional boxing of page `\thepage'}%
    \setbox\AtBeginShipoutBox=\hbox{\copy\AtBeginShipoutBox}%
  \fi
}

\usepackage{eso-pic}

\AddToShipoutPictureBG*{%
  \AtPageUpperLeft{%
    \hspace{0.75\paperwidth}%
    \raisebox{-3.5\baselineskip}{%
      \makebox[0pt][l]{\textnormal{DES 2019-0517}}
}}}%

\AddToShipoutPictureBG*{%
  \AtPageUpperLeft{%
    \hspace{0.75\paperwidth}%
    \raisebox{-4.5\baselineskip}{%
      \makebox[0pt][l]{\textnormal{FERMILAB-PUB-20-029-AE}}
}}}%

\usepackage{float}

\usepackage{graphicx}	

\title[OzDES multi-object fibre spectroscopy]{OzDES multi-object fibre spectroscopy for the Dark Energy Survey: Results and second data release}

\input{files/DES-2019-0517_author_list_v20200417.dat}

\date{Accepted 2020 May 04. Received 2020 April 22; in original form 2020 March 19}

\pubyear{2020}

\begin{document}
\label{firstpage}
\pagerange{\pageref{firstpage}--\pageref{lastpage}}
\maketitle

\begin{abstract}
We present a description of the Australian Dark Energy Survey (OzDES)
and summarise the results from its six years of operations. Using the
2dF fibre positioner and AAOmega spectrograph on the 3.9-m
Anglo-Australian Telescope, OzDES has monitored 771 AGN, classified
hundreds of supernovae, and obtained redshifts for thousands of
galaxies that hosted a transient within the 10 deep fields of the Dark
Energy Survey. We also present the second OzDES data release,
containing the redshifts of almost 30,000 sources, some as faint as
$r_{\mathrm AB}=24$ mag, and 375,000 individual spectra. These data,
in combination with the time-series photometry from the Dark Energy
Survey, will be used to measure the expansion history of the Universe
out to $z\sim1.2$ and the masses of hundreds of black holes out to
$z\sim4$. OzDES is a template for future surveys that combine
simultaneous monitoring of targets with wide-field imaging cameras and
wide-field multi-object spectrographs.

\end{abstract}




\section{Introduction}


Over the next ten years, surveys using the next generation of
multi-object spectroscopic facilities fed by rapidly configurable
fibre positioning systems will obtain the spectra of tens of millions
of sources. Facilities such as the Subaru Prime Focus Spectrograph
\citep[PFS,][]{Tamura2018} on the 8.2-m Subaru telescope, the Dark
Energy Spectroscopic Instrument
\citep[DESI,][]{DESI2016,Vargus-Magana2019} on the 4-m Mayall
telescope, and the 4-metre Multi-Object Spectroscopic Telescope
\citep[4MOST,][]{deJong2019} on the 4-m VISTA telescope are either
being built or on the cusp of entering operation. Other facilities,
such as Mauna Kea Spectroscopic Explorer \citep[MSE,][]{MSE2019} and
MANIFEST \citep{Lawrence2018} for the Giant Magellan Telescope, are
being planned. Over the same time frame, the Legacy Survey of Space
and Time \cite[LSST;][]{LSST2017} at the Rubin Observatory will start
imaging the entire southern sky multiple times in multiple pass-bands
over 10 years. The advent of the LSST in combination with these
spectroscopic facilities will allow contemporaneous imaging and
spectroscopic studies of tens of thousands of targets.

As a herald to these future surveys, OzDES\footnote{Australian (aka
  ``Oz'') Dark Energy Survey.} has used the 2dF fibre positioner on
the Anglo-Australian Telescope (AAT) to obtain spectra of thousands of
sources in the 10 deep fields of the Dark Energy Survey (DES) over the
6 years than DES ran. OzDES has two main scientific goals: i)
constraining the dark energy equation of state parameter using type Ia
supernovae (SNe Ia), and ii) measuring the mass of super-massive black
holes over a broad range of redshifts using AGN reverberation
mapping. Additionally, OzDES has obtained spectra for a number of
ancillary projects, such as improving and quantifying the accuracy of
redshifts obtained from broad-band photometry. In total, about 40\% of
the fibre hours not used for active supernovae, host galaxies and AGN
were used for these projects.

This paper presents results from the full six years of OzDES, and
describes the second OzDES data release\footnote{Available from Data Central:
  \url{https://datacentral.org.au}. Future releases
  will be announced on Data Central}. After giving a brief summary of
the principal scientific aims of OzDES in Sec.~\ref{sec:DES+OzDES}, we
provide an overview of OzDES operations for the last three years of
the survey. Throughout the paper we will use Y4, Y5, and Y6 to denote
the fourth, fifth and sixth years of OzDES.  The observing strategy
for the first three years of OzDES (Y1, Y2 and Y3) and the precursor
survey are discussed in two papers: \citet{Yuan2015} gives an overview
up the end of Y1 (2013), including the OzDES precursor survey, and
\citet{Childress2017} gives an overview up until the end of Y3
(2015). We will refer to these papers as Yu15 and Ch17 henceforth. The
reliability and completeness of OzDES redshifts are analysed in
Sec.~\ref{sec:redshifts}, and the data release is described in
Sec.~\ref{sec:datarelease}. In the sections that follow, we examine
various aspects of the survey, such as the frequency at which we were
able to target AGN, and how the signal-to-noise ratio behaves with
time for objects that have the longest integration times: the host
galaxies of transients. Before giving the summary of the paper in the
final section, we compare OzDES to TiDES, a OzDES-like survey using
the 4MOST facility \citep{Swann2019}.  All magnitudes listed in this
paper are measured using 2\arcsec\ diameter apertures and are on the
AB magnitude system.

\section{DES and OzDES}\label{sec:DES+OzDES}

The Dark Energy Survey (DES) was a six-year programme, ending in 2019,
using the DECam imager \citep{Flaugher2015} on the CTIO Blanco 4m
telescope in Chile. DES consisted of two surveys: a wide-field survey
covering 5,000 square degrees, and a deep survey consisting of 10
DECam fields located in several well-studied extra-galactic
fields. These 10 fields, the coordinates of which can be found in
Table 2 of \citet{DAndrea2018}, were targeted repeatedly in {\it griz}
over the course of the first five years of DES with a cadence of
approximately 6 days ,\citep{Diehl2018}, and consist of 8 relatively
shallow fields and 2 extra deep fields (known as X3 and C3). The
regular cadence facilitated the discovery and follow-up of transient
sources, such as supernovae (SNe), and the monitoring of variable
objects, such as AGN.

DES aims to place constraints on the dark energy equation of state
parameter and deviations from General Relativity using four
astronomical probes: weak gravitational lensing, galaxy clusters, the
large scale distribution of galaxies, and type Ia supernovae. Results
from small subset of the DES data set have been published in a number
of papers, the most recent of which are constraints that combine
supernovae, gravitational lensing, and galaxy clustering
\citep{DES2019b}.

DES data have been used in many studies covering a diverse range of
fields \citep{DES2016}. Notable examples include studies of bodies in
the outer solar system \citep[for example]{Khain2018}, AGN accretion
disk sizes \citep{Mudd2018,Yu2020}, dwarf galaxies and stellar streams
in the halo of the Milky Way \citep[for example]{Marshall2019,Li2019}
and rare transients such as merging neutron stars \citep{Palmese2017},
super-luminous supernovae \citep{Smith2018,Angus2019}, and rapidly
evolving transients \citep{Pursiainen2018}.

Since 2012, OzDES has been targeting sources in the 10 DES deep fields
with the 2dF fibre positioner on the AAT. As shown in Fig.~1 of
\citet{Yuan2015}, the fields of view of 2dF and DECam are very
similar. In addition to the 10 DES deep fields, OzDES targeted the DES
MaxVis field during one of the OzDES observing runs. Centred at
RA=$97.5\degree$ and Dec=$-58.75\degree$, it is observable during the
entire DES observing season. It is one of the DES standard star fields
that was observed during twilight. \cite{Yu2020} used quasars in this
field to study continuum reverberation mapping.

The scientific aims of OzDES focus on two areas: i) using redshifts
and distances of type Ia supernovae to measure the expansion history
of the Universe and thereby constrain the dark energy equation of
state parameter; and ii) measuring the time lags between the continuum
and the broad lines in a sample of AGN to measure their central black
hole masses and to investigate the possibility of using AGN as a
luminosity distance indicator that would explore a redshift range that
is currently beyond the reach of supernovae.

Over the six years that OzDES ran, OzDES obtained redshifts for 7,000
candidate supernova hosts \citep{DAndrea2018} and spectroscopically
confirmed several hundred supernovae (SNe). These redshifts will be
used with distances to the supernovae inferred from broad-band
lightcurves obtained by DES to measure the expansion history of the
Universe and to constrain the dark energy equation of state
parameter. Using an initial sample of 207 spectroscopically confirmed
type Ia supernovae from the first three years of data in combination
with 122 low redshift supernova from the literature and constraints
from the Cosmic Microwave Background (CMB), the \citet{DES2019a}
obtained a 7\% constraint on a constant dark energy equation-of-state
parameter. Analysis of the full DES sample is ongoing, and the final
cosmological results will include approximately an order of magnitude
more supernovae, the majority of which will be classified
photometrically.

At the same time, OzDES has monitored a sample of 771 AGN up to
$z\sim4$ with the goal of measuring black hole masses using the time
lag between variations in the continuum, measured from the broad-band
photometry obtained by DES, and the broad lines, measured from the
spectra obtained by OzDES. The first time lags, based on the first
four years of data, were published in \citet{Hoormann2019}.

In addition to these two main projects, OzDES is facilitating the
studies of a number of areas, both within DES and outside of DES, by
obtaining redshifts of targets of interest in these 10 fields using
the ``spare" fibres that are not allocated to an AGN, an active
transient, or a host galaxy. The most notable examples are: radio
galaxies from the ATLAS survey \citep{Franzen2015}; brightest cluster
galaxies \citep{Webb2015}; redMaGiC galaxies
\citep{Rozo2016,Sanchez2018}; and luminous red galaxies (LRGs), which
are used in combination with redshifts from other sources to train
photometric redshift algorithms
\citep{Sanchez2014,Bonnett2016,Gschwend2018}.

\section{OzDES Operations for Y4, Y5 and Y6}\label{sec:operations}

Starting in 2013, OzDES ran for six observing seasons, with an initial
allocation of 12 nights in Y1 (AAT semester 2013B), and allocations of
16, 20, 20, 20, and 12 nights in subsequent years. In 2012, when
aspects of the DES survey strategy were being tested during a period
of science verification (SV), 5 nights were allocated to an OzDES
precursor survey. As for DES, these 5 nights were used to develop and
test the observing strategy that would be used for the next six years.

Observations typically started in late August or early September and
terminated four to five months later in December or January the
following year. The observing log for Y4, Y5, and Y6 can be found in
Appendix A and follows the format presented for earlier years in Yu15
and Ch17.

In total, there were 41 numbered runs, with the last night of run 41
occurring on 09 January 2019. Some of the nights during these runs
were shared with other programmes, such as 2dFLens \citep{Johnson2017}
during Y2 and Y3, XXL \citep{Pierre2016} in Y4, DEVILS
\citep{Davies2018} in Y5, and the HSC Transient Survey
\citep{Yasuda2019} in Y6. During Y2 and Y3, OzDES and 2dFLens shared a
common run numbering scheme, and there are a number of runs during
those years in which only 2dFLens fields were observed.

The end of Y5 marked the end of the DES programme to discover transients
in the 10 DES deep fields. DES continued to target fields in the wide
survey during Y6. OzDES also continued into a sixth year, but was no
longer targeting transients. Instead, OzDES continued with its programme
to monitor AGN \citep{King2015,Hoormann2019} and to obtain redshifts
of galaxies that hosted transients in earlier years. DES continued to
target the 10 DES supernova fields during Y6, but with a reduced
frequency of about once a month.

These data are used to spectrophotometrically calibrate the spectra of
the AGN in the deep fields that were being monitored by OzDES.


\subsection{Instrumental Setup}

Apart from the final OzDES run in Y6 (run 041), the instrumental setup
used in Y4, Y5 and Y6 was identical to the set up used in Y3. The set
up in Y1 and Y2 and the precursor survey were slightly
different. During these years, the central wavelength of the blue
grating was set 20\AA\ to the blue. Details of the setup and the
reasons for the change between Y2 and Y3 can be found in Ch17.

The final OzDES run was not officially scheduled as an OzDES run. At
the beginning of 2019, an opportunity arose to take additional data in
the C3 and X3 fields during time that was allocated to the HSC
Transient Survey, which has similar aims to the SN programme in OzDES,
but is targeting fainter, more distant host galaxies. For this run,
and this run only, we used the setup used by the HSC Transient Survey,
which uses the x6700 dichroic (OzDES uses the x5700 dichroic) and a
wavelength setting that results in a wavelength coverage that is
shifted about 1000\AA\ to the red, from 3700\AA--8800\AA\ to
4700\AA--9800\AA.

\subsection{OzDES Target Allocation}\label{sec:allocation}

OzDES targeted a wide range of sources over the six years it ran, with
active transients, AGN, and host galaxies with $m_r<24$ having the
highest priority and occupying most of the fibres. A full listing of
the source types that were observed in the last three years, together
with their priorities, is provided in Table~\ref{tab:sourcetypes}.

In addition to a priority, each type had a quota. Starting with the
highest priority (priority 8), targets were randomly selected until
their was no more targets to select or the quota for that type of
source had been reached. Further details on the target selection
algorithm are described in Y15.

\begin{table*}
 \caption{A listing of the source types observed during Y4, Y5, and
   Y6, together with the nominal priority. The list is ordered in
   priority from highest (8) to lowest (1). Also listed are source
   types (noted with asterisks) from the first three yeas of OzDES
   that were discontinued before the start of Y3. Source types for the
   MaxVis field are listed in Appendix \protect{\ref{sec:MaxVis}}.}
 \label{tab:sourcetypes}
 \begin{tabular}{lll}
  \hline
  Type & Priority & Comment \\
  \hline
  Transient & 8 & Active transients \\
  AGN & 7 & Reverberation mapping programme \citep{Hoormann2019}\\
  SN\_host$^a$ & 6 & Hosts with $m_r < 24$ \\
  Cooke\_host & 6 & Host galaxy candidates of transients found in deep SNLS$^b$ stacks\\
  StrongLens$^*$ & 6 & Candidate strong lenses \citep{Nord2016,Jacobs2019}\\
  White Dwarfs$^*$ & 6 & White dwarfs to aid calibration\\
  DEVILS$^c$  & 5 & Objects in the DEVILS survey \citep{Davies2018} \\
  SNLS  & 5 & SN hosts from the SNLS survey data \citep{Astier2006,Betoule2014}\\
  Tertiary$^*$  & 5 & Stars with a broad range of colours to aid calibration\\
  Cluster Galaxies  & 4 & Cluster Galaxies (some targets were observed at higher priority) \\
  Radio Galaxies  & 4 & Radio Galaxies (some targets were observed at higher priority) \\
  QSO  & 4 & Faint QSOs in the S1 and S2 fields\\
  XXL\_QSO  & 4 & QSOs in the XXL fields \citep{Pierre2016}\\
  SN\_host\_faint & 4 & Hosts with $24 < m_r < 25$\\
  RedMaGiC  & 4 & For calibrating photometric redshifts of redMaGiC galaxies \citep{Rozo2016} \\
  SpARCS$^d$ & 4 & Brightest Cluster Galaxies from the SpARCS survey  \citep{Wilson2009}\\
  ELG$^*$  & 4 & Emission Line Galaxies \\ 
  SN\_free\_host & 3 & Host galaxies of previously targeted transients\\
  Photo-$z$ & 3 & For training and testing photometric redshift algorithms \\
  LRG  & 2 & Luminous Red Galaxies\\
  Bright galaxies  & 1 & Back up targets during poor conditions\\
  Bright stars$^*$  & 1 & Back up targets during poor conditions\\
    \hline \\
    \multicolumn{3}{l}{\footnotesize$^a$ The type "SN host" not only includes galaxies that hosted a SN, but also galaxies that hosted transients that were not SNe.}\\
    \multicolumn{3}{l}{\footnotesize$^b$ SNLS - Supernova Legacy Survey}\\
    \multicolumn{3}{l}{\footnotesize$^c$ DEVILS - Deep Extragalactic Visible Legacy Survey}\\
    \multicolumn{3}{l}{\footnotesize$^d$ SpARCS - Spitzer Adaptation of the Red-Sequence Cluster Survey}\\
   
 \end{tabular}
\end{table*}

If conditions were poor because of low transparency (generally more
than 2 mag of extinction) and/or poor seeing (generally poorer than 3
arc-seconds) we switched to a backup programme that mostly consisted of
bright galaxies. However, transients and AGN in the RM programme were
always targeted.

Over the seasons, the composition of sources and their priorities
changed. The biggest changes occurred at the end of Y5 and the
beginning of Y6. In Y6, no new transients were being discovered, as
DES stopped monitoring the 10 deep fields for transients; however,
during Y6, we did continue to observe a small number of transients
that were discovered in earlier years.

There were other changes too. At the end of Y5, the quota for SN hosts
increased from 100 (200 for X3 and C3, the two extra deep fields) in
the first five years to 200 (250 for X3 and C3), and two new classes
of sources were added. Given the success at obtaining redshifts for
sources as faint as $r_{\mathrm AB}\sim 24$ and no strong evidence to
a floor in the noise (Ch17), we added galaxies with $24 < r_{\mathrm
  AB} < 25$. We labelled these as {\tt SN\_host\_faint}. We had
targeted hosts as faint as this in past seasons, but the aim then was
different. Then, if we did not see an emission line that enabled us to
get a redshift after one run, we discarded the source. During Y6, we
continued observing these faint hosts until a secure redshift (see
Sec.~\ref{sec:redshifts}) was obtained.

We also added host galaxies of transients that were previously
observed when the transient was still bright enough to contribute to
the flux density.  We labelled these as {\tt SN\_free\_host}. We will
use these spectra to examine the properties of host galaxies without
light from the supernova, and to subtract the galaxy spectrum from
earlier spectra that contained light from the both the galaxy and the
transient. The aim here is to see if one can identify the type of
transient in spectra where this was not possible. This has been proven
to work for spectra taken with slits \citep{Dawson2009}, but has never
been tried for spectra obtained with fibres.

\begin{figure}
 \includegraphics[width=\columnwidth]{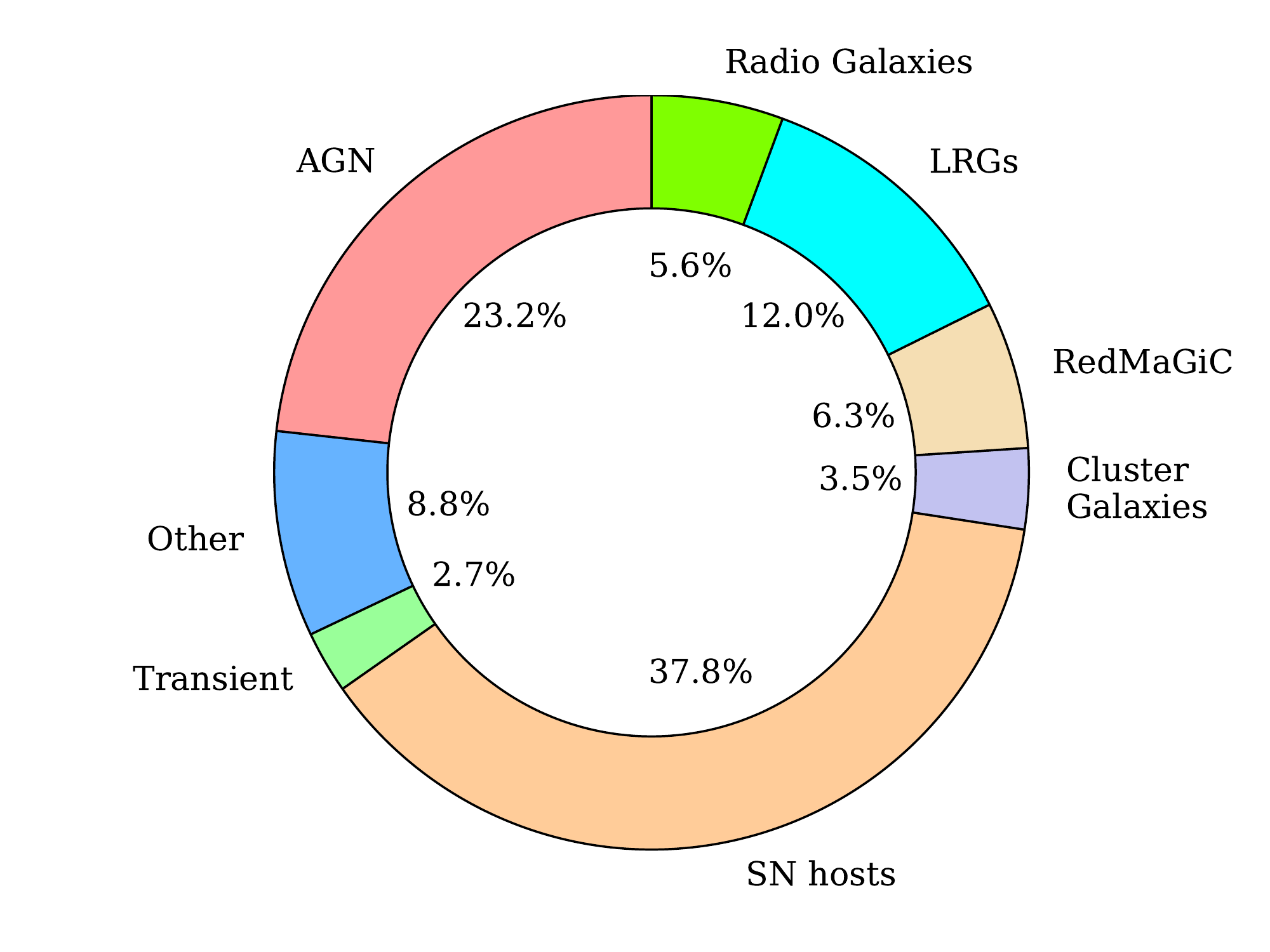}
 \includegraphics[width=\columnwidth]{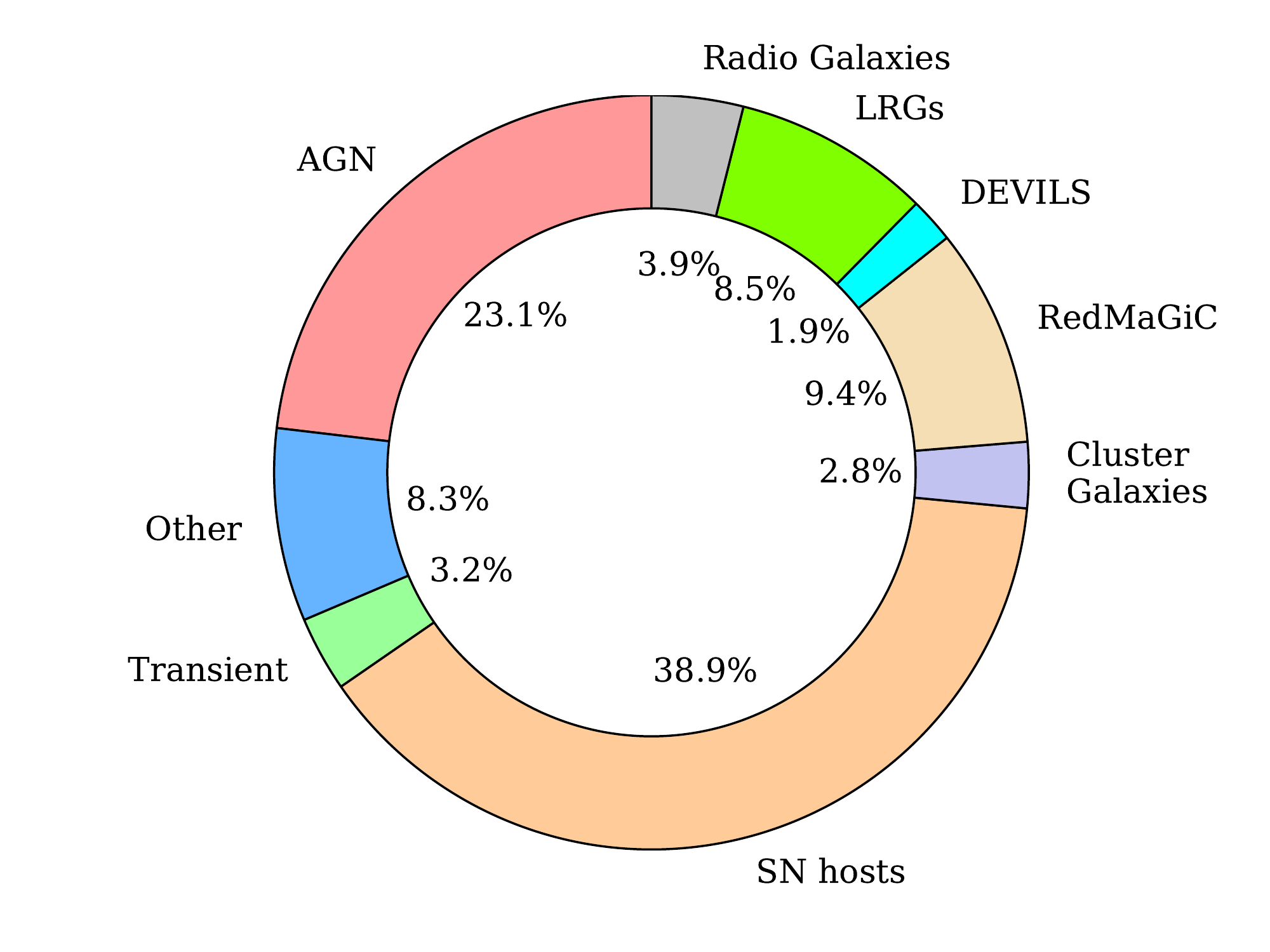}
 \includegraphics[width=\columnwidth]{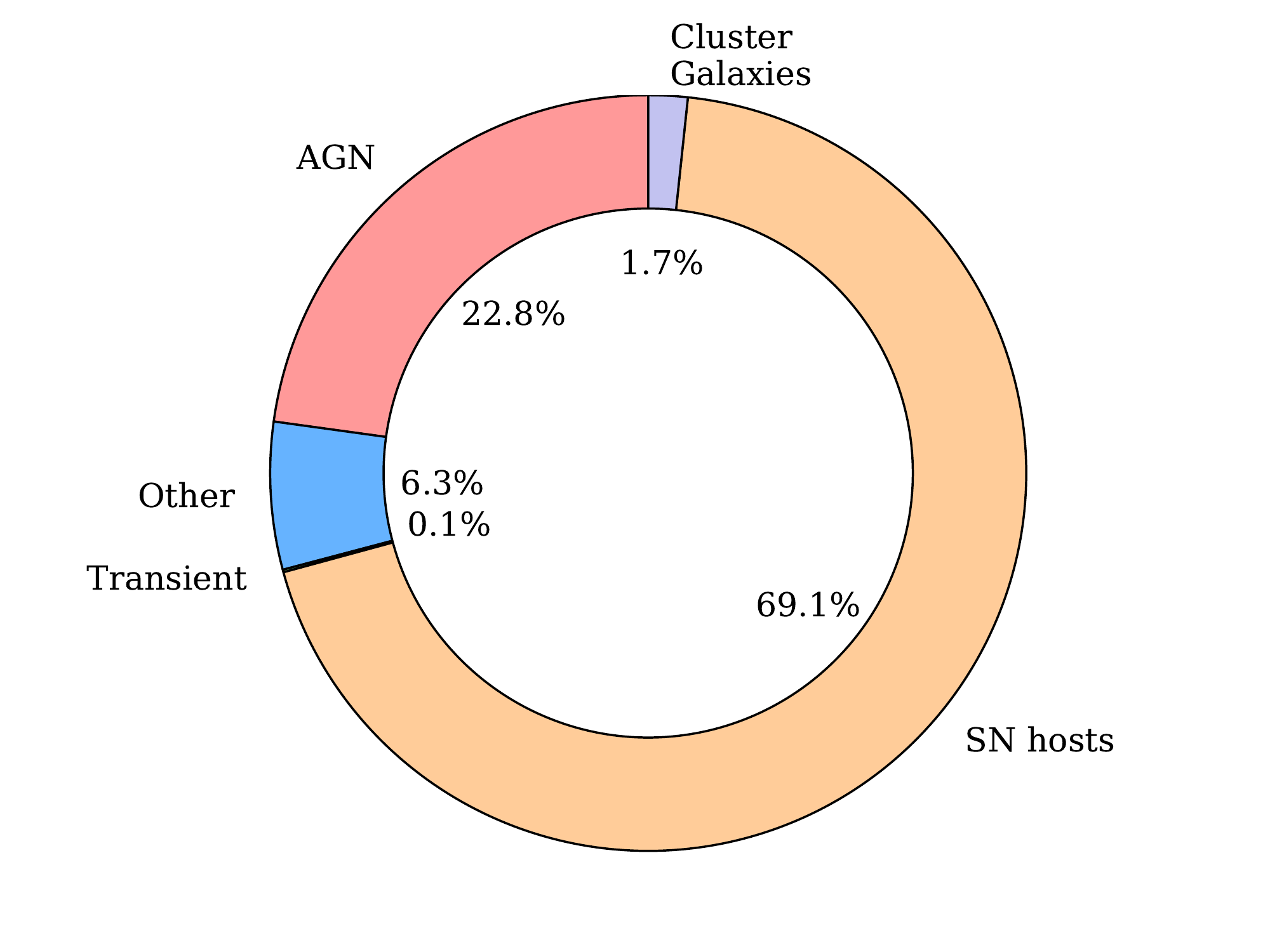}

 \caption{Plots showing how the allocation changed in the last three
   years of OzDES: Y4 at the top, Y5 in the middle, and Y6 at the
   bottom. The percentages represent the fraction of fibre hours. In
   the final year almost all targets were either AGN or SN hosts.}
 \label{fig:allocations}
\end{figure}

Fibre allocations for the last three years of OzDES are listed in
Table~\ref{tab:allocations} and shown as plots in
Fig~\ref{fig:allocations}. A similar table in for the first three
years is provided in Ch17. After Y1, a quarter of all fibres were
allocated to AGN that were being followed in the OzDES reverberation
mapping. The number of fibres allocated to SN hosts steadily increased
from 12\% in Y1 to 39\% in Y5. The large jump from Y5 to Y6 in the
number of SN hosts reflects the increased quota for SN hosts in Y6 and
the addition of two new SN host categories. Almost all fibres in Y6
were allocated to an AGN or a SN host.

\begin{figure}
 \includegraphics[width=\columnwidth]{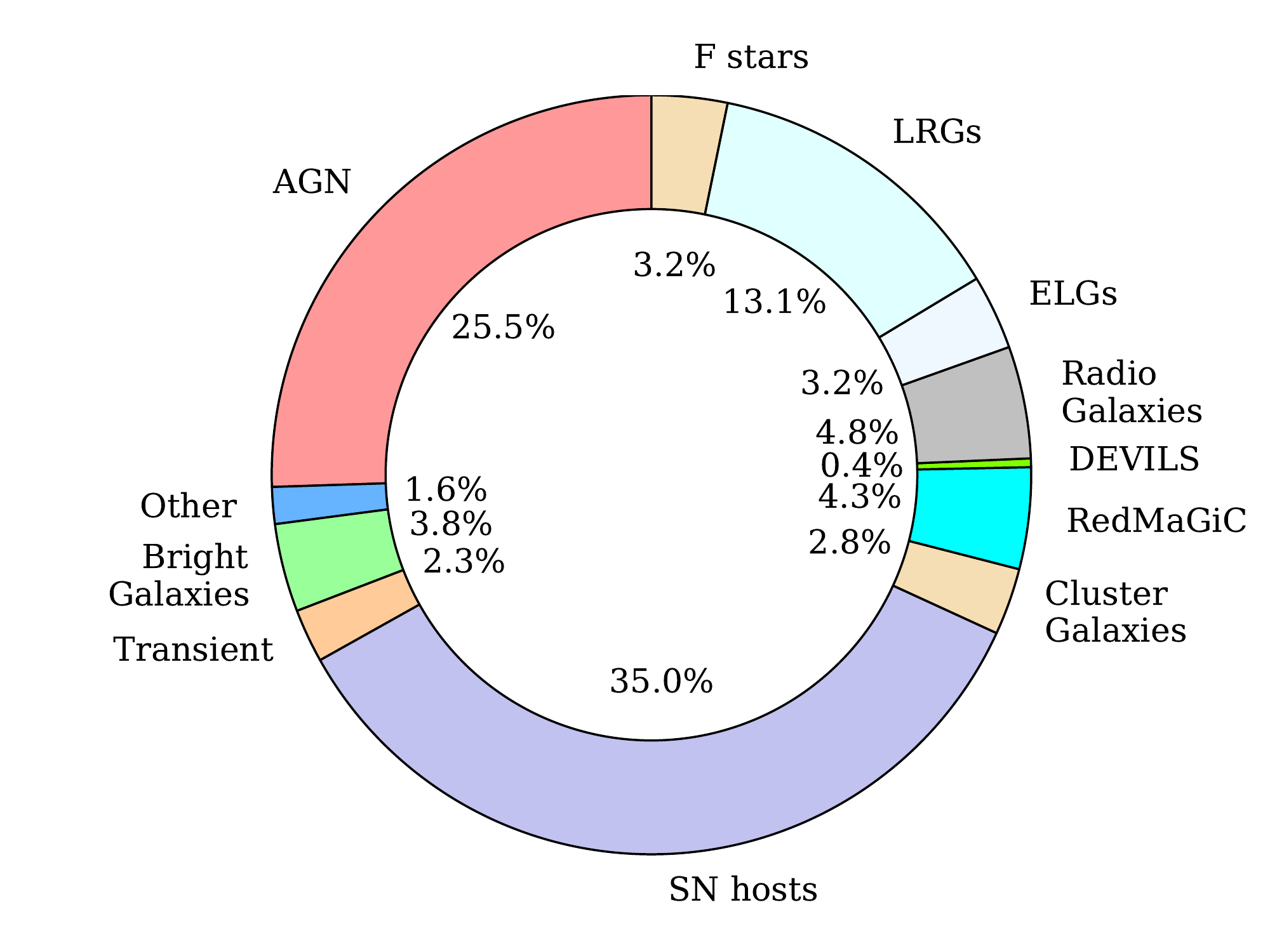}

 \caption{Plot showing the relative allocation for the full six years
   of OzDES. The percentages represent the fraction of fibre hours.}
 \label{fig:fullallocations}
\end{figure}

Fibre allocations over the full OzDES survey are shown in
Fig~\ref{fig:fullallocations}. The three most commonly observed
sources were SN hosts (35\%), AGN (26\%) and galaxies used to train
photometric redshift algorithms: LRGs, ELGs, and redMaGiC galaxies
(21\%).

There can be considerable overlap in the individual target catalogues
that make up the OzDES target input catalogue. For example, a SN host
can also be listed as a radio galaxy, LRG, or redMaGiC galaxy. In
Table~\ref{tab:allocations}, Fig~\ref{fig:allocations}, and
Fig~\ref{fig:fullallocations} objects retain the type that has the
highest priority. For example, a source that is listed in the OzDES
target input catalogue as a SN host, radio galaxy, an LRG, and a
redMaGiC galaxy will be classified and counted as a SN host.

\begin{table*}
\begin{center}
 \caption{Fibre allocations for the last three years of OzDES (Y4, Y5, and Y6)}
 \label{tab:allocations}
 \begin{tabular}{lrrrrrr}
  \hline
   & \multicolumn{2}{c}{Y4} & \multicolumn{2}{c}{Y5} & \multicolumn{2}{c}{Y6}\\
   Object & Fibre & Fraction of  & Fibre & Fraction of & Fibre & Fraction of\\
          & Hours & Fibre Hours & Hours & Fibre Hours & Hours & Fibre Hours\\

  \hline
          Transients &    939 &   2.7\%  &   1195 &   3.2\% &     29 &   0.1\%  \\
                 AGN &   8191 &  23.2\%  &   8578 &  23.1\% &   6115 &  22.8\%  \\
            SN hosts &  13333 &  37.8\%  &  14413 &  38.9\%  &  18513 &  69.1\%  \\
    Cluster Galaxies &   1219 &   3.5\%  &   1048 &   2.8\% &    449 &   1.7\%  \\
      Radio Galaxies &   1989 &   5.6\%  &   1455 &   3.9\% &      0 &   0.0\%  \\
    	      DEVILS &      0 &   0.0\%  &    720 &   1.9\% &      0 &   0.0\%  \\
            RedMaGiC &   2220 &   6.3\%  &  3481 &   9.4\% &      0 &   0.0\%  \\
                LRGs &   4235 &  12.0\%  &   3137 &   8.5\% &      0 &   0.0\%  \\
               Other &   3107 &   8.8\%  &   3071 &   8.3\% &   1688 &   6.3\%  \\
          \bf{Total} &  35232 &          &  37097 &         &  26793 &          \\

   \hline
 \end{tabular}
\end{center}
\end{table*}

\begin{table}
 \caption{Fibre allocations for the entire OzDES survey}
 \label{tab:allYears}
 \begin{tabular}{lrr}
  \hline
   & \multicolumn{2}{c}{All years} \\
   Object & Fibre & Fraction of  \\
          & Hours & Fibre Hours \\
          \hline
          Transients &   4438 &   2.3\% \\
                 AGN &  49548 &  25.5\% \\
            SN hosts &  68028 &  35.0\% \\
    Cluster Galaxies &   5523 &   2.8\% \\
              DEVILS &    720 &   0.4\% \\
      Radio Galaxies &   9271 &   4.8\% \\
            RedMaGiC &   8394 &   4.3\% \\
             F stars &   6247 &   3.2\% \\
                LRGs &  25521 &  13.1\% \\
                ELGs &   6175 &   3.2\% \\
               Other &   3043 &   1.6\% \\
     Bright Galaxies &   7298 &   3.8\% \\
          \bf{Total} & 194207 &\\
          
  \hline
 \end{tabular}
\end{table}
  
\subsection{Observing Strategy, Calibration and Data Reduction}

During each run, we aimed to get at least two 2400 second exposures
per field. This was not always possible due to the length of some runs
and weather conditions. Of the 31 runs that were scheduled for OzDES,
we were able to observe all 10 fields on 16 occasions. There were only
two, relatively short runs where we did not observe any fields because
of poor weather.

Overall, $\sim25$\% of time was lost to poor observing
conditions. Another $\sim5$\% of the time was sub-optimal (thick
cirrus and/or poor seeing), and was used for the back up
programme. Observing logs for the final three years of OzDES, including
the amount of time lost for poor weather, are included in the
appendixes.

\begin{figure}
 \includegraphics[width=\columnwidth]{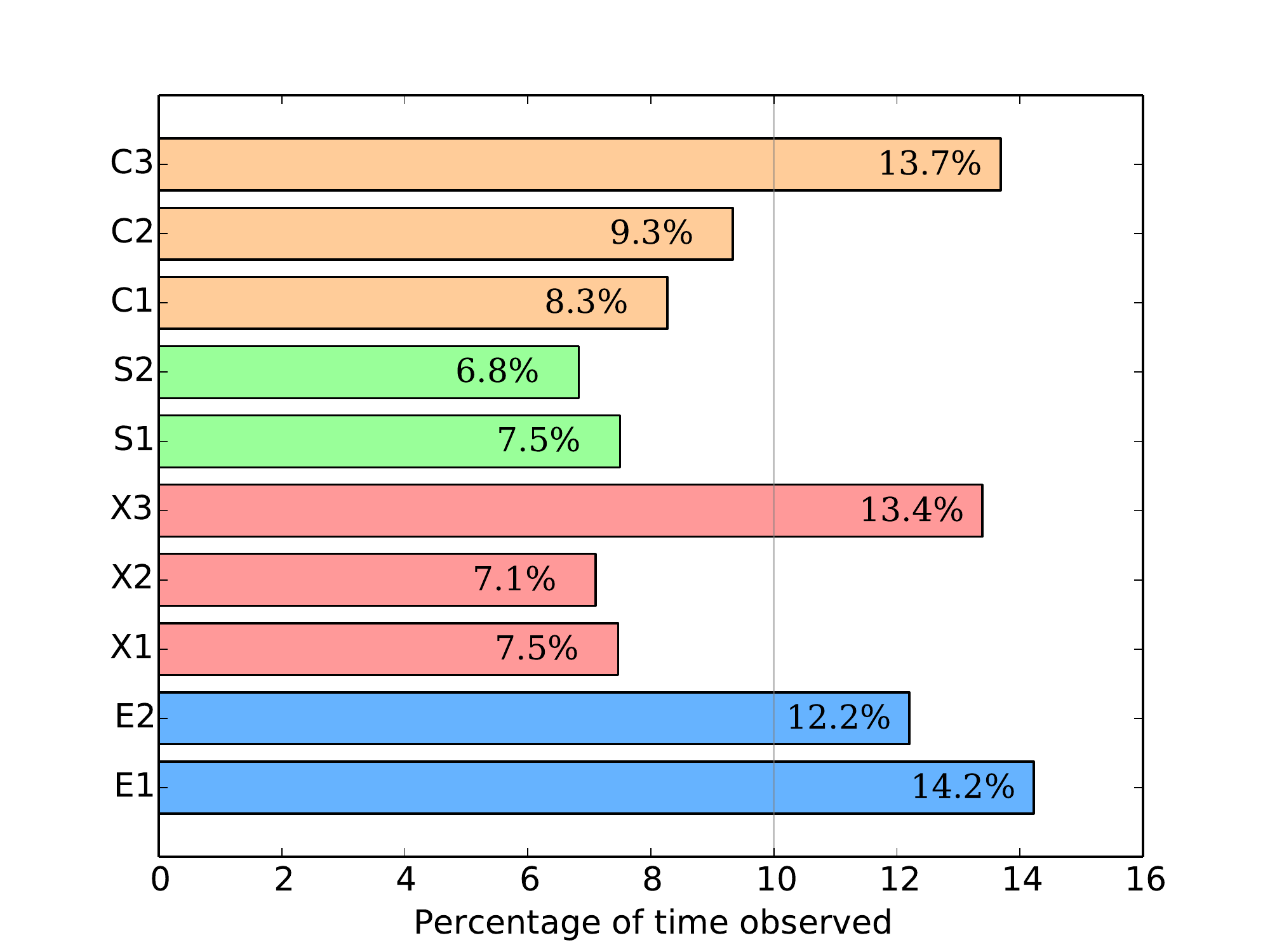}
 
 \caption{Bar plot showing the percentage of time each field was
   observed. The vertical line sits at 10\%, which is where all fields
   would land if they had been observed with the same frequency. The X3,
   C3, E1, and E2 fields were the ones most frequently observed,
   reflecting the priorities these fields had and their locations in
   the sky.}
 \label{fig:frequency}
\end{figure}

The two extra deep fields, C3 and X3, had the highest priority, so
these fields were the most frequently observed, as can be seen in
Fig.~\ref{fig:frequency}. The two E fields, E1 and E2 were also
observed frequently, since these fields were often the only ones
visible at the beginning of the night during the start of the
observing season and are the most southern of the ten deep fields,
which results in a longer window of observability from the AAT. Next
in priority were the six remaining fields: X1, X2, C1, C2, S1, and S2.

In addition to the sources listed in Table~\ref{tab:sourcetypes},
about a dozen fibres per field were allocated to F stars, and another
25 ``sky'' fibres were placed on positions that were free of
sources. The targeting priority for both the F stars and the ``sky''
fibres was set to 5, which is below that of Transients, AGN and SN
hosts, but higher than other types of objects. See
Table~\ref{tab:sourcetypes} for details.

\begin{figure}
 \includegraphics[width=\columnwidth]{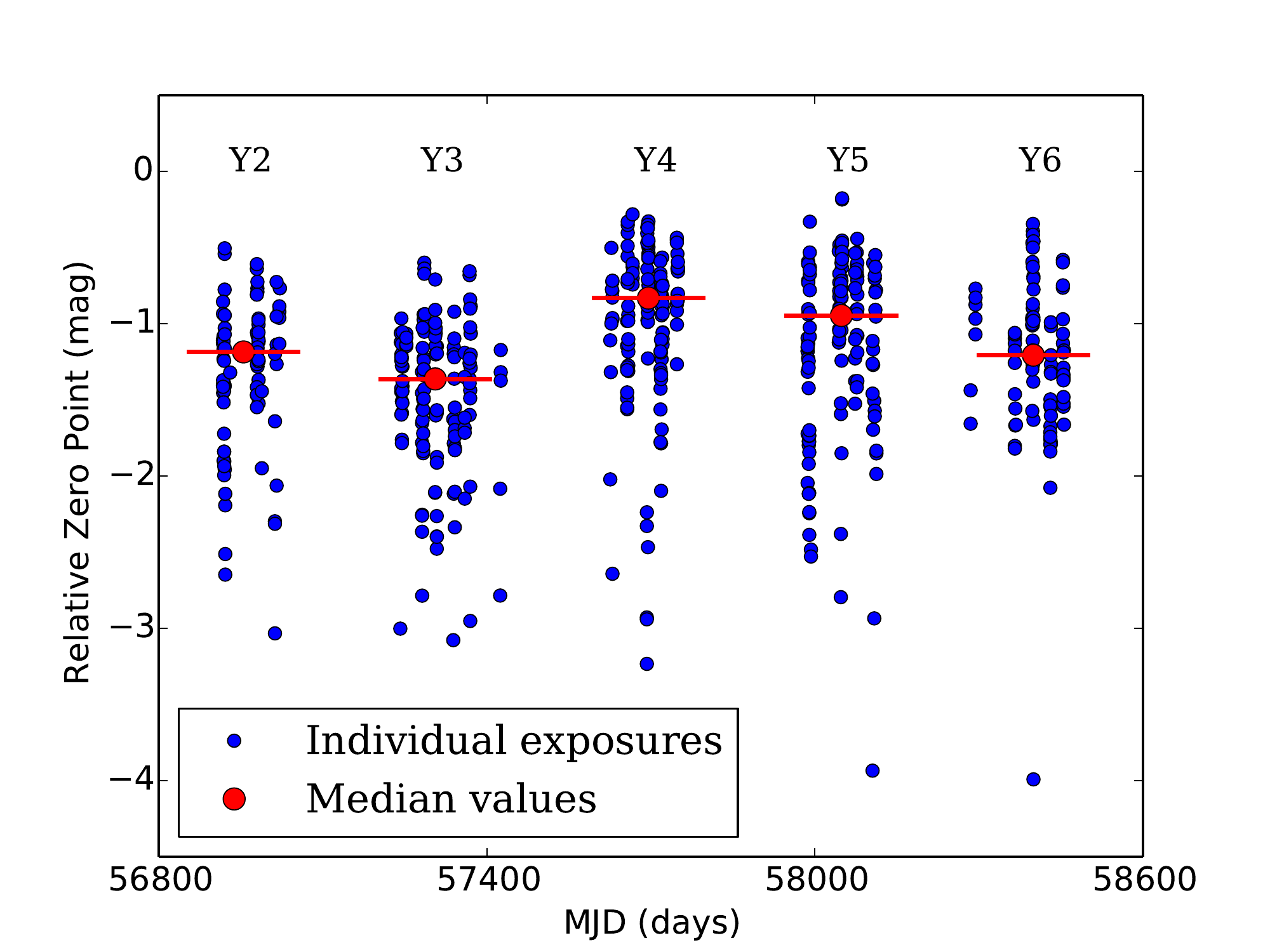}
 
 \caption{Plot of the relative throughput (zero point) of data taken
   from Y2 to Y6. The median value for each year is shown as the red
   dot and the red line. The scale is in magnitudes and more positive
   numbers indicate higher throughput. In Y1, very few F-stars were
   observed. Consequently there are no reliable throughput
   measurements for the first year of OzDES. The relative zero point
   plotted here differs from the zero point recorded in the FITS
   headers by an additive constant of 33. See footnote d to
   \protect{Table~\ref{tab:keywords}} for details on how the zero point is
   computed.}
 \label{fig:throughput}
\end{figure}

The F stars were primarily used to estimate throughput. The throughput
was measured by integrating the extracted and wavelength calibrated
spectra though the DECam filter band passes \citep{Flaugher2015}. For the red arm, we used
the DECam $r$ filter: for the blue arm, we used the DECam $g$ filter. The
relative throughput of the red arm over the course of OzDES is shown
in Fig.~\ref{fig:throughput}. The throughput is reported in
magnitudes, so each tick on the vertical access corresponds to a
difference of about 2.5 in throughput.  The vertical scatter is
largely driven by transparency and atmospheric seeing.

Spectra that were taken when the throughput was excessively low (a
couple of magnitudes lower than the medians shown in
Fig.~\ref{fig:throughput}) are not used in the coadded spectra;
however, they are included in the data release. For such spectra, the
QC keyword in the FITS header is given the value {\tt poorConditions.}

In the early years, we used the F-stars to derive the transfer
functions for the red and blue arms that were then used throughout the
survey, and to develop techniques that used the broadband DES
photometry to spectrophotometrically calibrate the spectra of AGN used
for the OzDES RM project \citep{Hoormann2019}.

All data were processed with the OzDES pipeline, which uses a modified
copy of v6.46 of 2dfdr\footnote{\url{https://aat.anu.edu.au/science/software/2dfdr}}
\citep{Croom2004} and our own bespoke python scripts. The steps are
described in Yu15 and Ch17, so we do not repeat them here. Since Ch17,
a small number of improvements to the pipeline have been made, the
most significant of which were a more robust algorithm to determine
the factor that scales spectra in the blue and red arms, modifications
to the PCA sky subtraction algorithm to make it less sensitive to
cosmic rays that were not fully masked, and more complete masking of
artifacts. These artifacts are described in section
\ref{sec:dataquality}.

All data were processed at the telescope, and combined with data taken
during earlier observing runs. Once processed, all raw and reduced
files were then archived on a server in Canberra. Typically, this was
done daily, after the night had ended, but before the start of the
next night. After each field was processed, the data was made
available for OzDES team members\footnote{Colloquially referred to as
  ``redshifters.''} dedicated to the task of measuring redshifts
within 24 to 48 hours of the data being taken. With the exception of
transients and AGN, sources with secure redshifts (see the next
section) were removed from the target catalogues and not re-observed
during subsequent nights.  During some of the longer OzDES runs, we
observed a number of the DES fields on multiple occasions. If this
occurred, then all objects that were observed on the first occasion
would be given higher priority in subsequent observations, unless they
had a secure redshift and had been removed from the target catalogue.
We have reprocessed all the data with the same version of the pipeline
on a couple of occasions. This ensures that all data are processed in
a consistent manner. The latest version of the OzDES pipeline is
version 18.21, and all data presented in this paper have been
processed with that version.

\section{Redshifts}\label{sec:redshifts}

As noted in Ch17, redshifts have been measured using {\tt MARZ}
\citep{Hinton2016} since the start of Y3. Earlier than this {\tt
  RUNZ}, which was developed by Will Sutherland, had been
used. Measuring redshifts is a task that is performed by two
redshifters, whose results are collated, scrutinised and then merged
by the chief redshift whip. Merging is also done using {\tt MARZ}. The
merged results then enter the OzDES database, from which all other
products are derived.

OzDES uses a redshift flag to quantify the confidence that the
reported redshift is correct. As reported in Yu15 and Ch17, the flag
can take on five main values.

\begin{itemize}
\item $Q = 4$, redshift based on multiple strong spectroscopic features matched, $> 99$\% confidence.
\item $Q = 3$, redshift based typically on a single strong spectroscopic feature or multiple weak features, $> 95$ \% confidence.
\item $Q = 2$, potential redshift associated with typically a single weak feature, low confidence, not to be used for science.
\item $Q = 1$, no matching features, thus no constraints on redshift. 
\item $Q = 6$, securely classified star.
\end{itemize}

As the survey progressed, we monitored the growing redshift catalogue
for sources that had conflicting redshifts. These are sources in which
the redshift changes between observing runs but the redshift quality
flag, $Q$, remains unchanged and $Q>2$. An example is a source that is
assigned a $Q=3$ redshift during one run, but with the addition of
more data in a later run, is assigned a different redshift ($\Delta z
> 0.05$) with the same quality flag.

On two occasions, once at the end of Y4 once at the end of Y6 , we
collated all cases where there was conflict, defined on these
occasions having a redshift difference larger than 0.05, and
reprocessed and re-redshifted these sources in the same way we
redshifted sources during each observing run. The revised redshifts
were then added to our redshift catalog. At the end of Y4, there were
a total of 191 sources that had conflicting redshifts. At the end Y6,
there were a total of 49, perhaps illustrating a degree of improvement
by the human redshifters.

In preparation for the second OzDES data release, we revisited cases
with conflicting redshifts, this time decreasing the threshold from
0.05 to 0.01 in redshift. We also included sources that were
classified with $Q=3$ quality redshift at some stage during the survey
but ended up with a lower redshift quality flag once all data were
combined. Combined, about 820 sources were re-examined.

\subsection{Redshift Reliability}

Using data taken up until the end of Y1, Yu15 estimated the
reliability of redshifts that are assigned $Q=3$ and $Q=4$ using
external catalogues and repeat observations of the same target. They
find that $Q=4$ quality redshifts are 100\% reliable and $Q=3$ quality
redshifts are $\sim 90\%$ reliable. After Y1, a number of measures
were put in place to lift the reliability of the $Q=3$ quality
redshifts. Our target for $Q=3$ quality redshifts at the start of the
OzDES survey was $\sim 95\%$. As discussed below, we have reached this
goal.

Ch17 used a different approach to assessing redshift reliability, 
taking advantage of the observing
strategy applied to SN hosts. Unlike other target types, SN hosts were
not removed from the input target catalogues until they attained a
$Q=4$ quality redshift. For those SN hosts with $Q=4$ quality
redshifts that initially attained a $Q=3$ quality redshift, one can
compare the initial redshift with the final redshift and get an
estimate of reliability of the $Q=3$ quality redshifts, assuming that
the $Q=4$ quality redshifts have a known reliability, which we take to
be 100\%, following that found in Yu15.

\begin{figure}
 \includegraphics[width=\columnwidth]{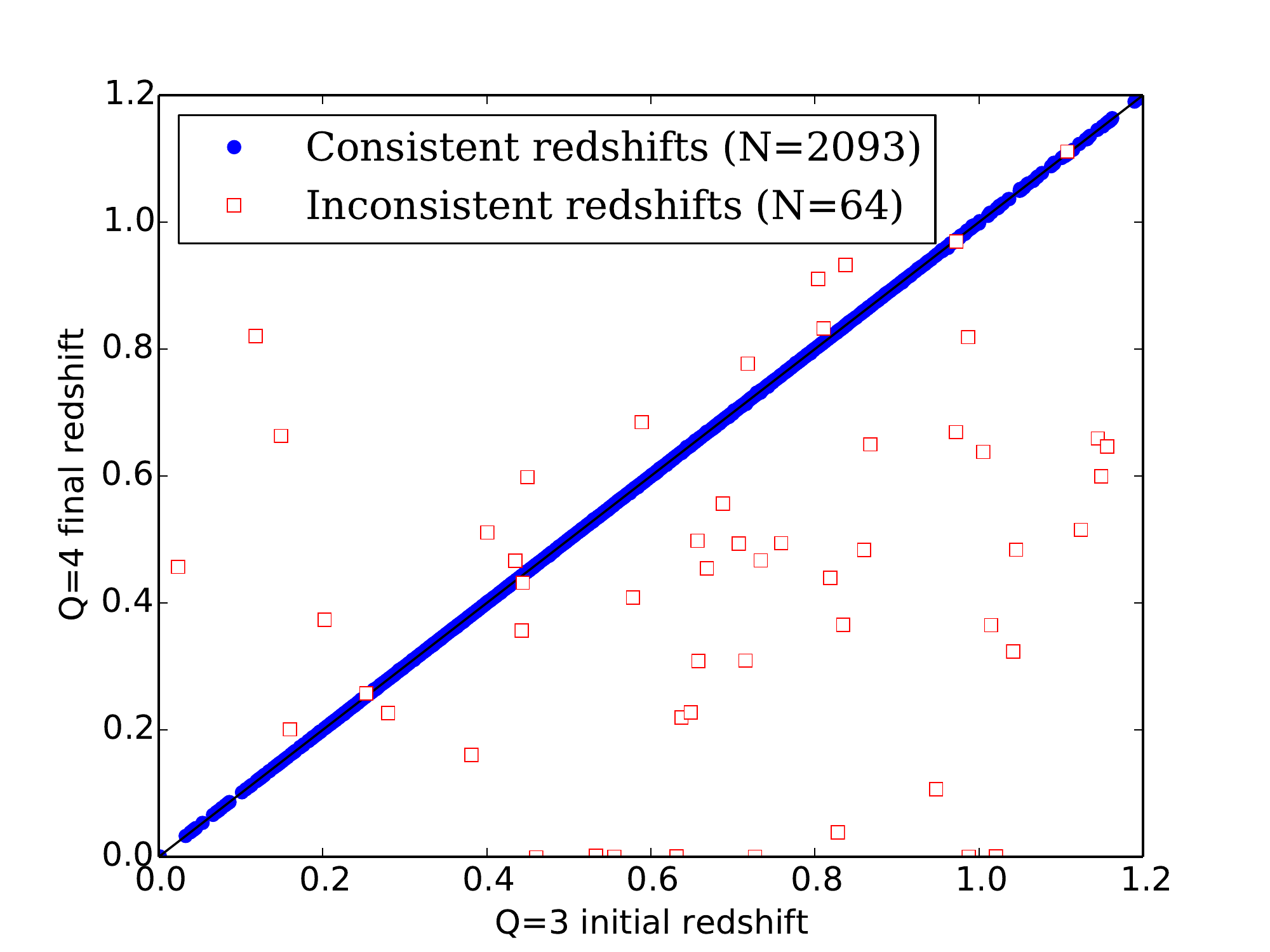}
 
 \caption{Redshift comparison for OzDES SN hosts that attained a $Q=3$
   quality redshift before attaining a $Q = 4$ quality redshift after
   further observations. Outliers (defined as $\Delta z > 0.003$) are
   shown as red squares. Sources with concordant redshifts are shown
   as blue points. The black line represents the one-to-one
   relationship. Only 3.1\% of sources changed redshifts when moving from $Q=3$ to $Q=4$.}
 \label{fig:reliability}
\end{figure}

Using data up until the end of Y3, Ch17 found that the redshifts of 17
out 443 sources (3.8\%) changed between the initial $Q=3$ quality
redshift and the final $Q=4$ quality redshift. We repeated this
analysis, using a sample that is almost five times larger.  We find that 64
of 2093 sources (3.1\%) changed redshifts, a slightly smaller fraction
than found in Ch17. The results are shown in Fig.~\ref{fig:reliability}.

The SN hosts allow one to examine the reliability of the $Q=3$ quality
redshifts in a second, complimentary way. A number of the fainter SN
hosts have been observed over several years and 20 or more observing
runs. As more data were obtained and added to earlier data, they would
have been re-redshifted, sometimes twice or more per run, which can
occur if they were observed over multiple times during a run. Some
sources have dozens of redshift measurements. Every time a source was
redshifted, there is a chance of an erroneous redshift being assigned
to that source. We can get a rough estimate of how often this happens
by searching for objects that obtained a $Q=3$ quality redshift during
one run, but did not subsequently obtain a $Q=3$ or $Q=4$ quality
redshift with more data.

We search the database for examples where a $Q=3$ quality redshift was
assigned once and not assigned again, even if the exposure time had
increased by a factor of two or three by the end of Y6. For these two
cases, we identified 31 and 11 SN hosts, respectively. For these
sources, we reduce the redshift quality flag from $Q=3$ to
$Q=2$. However, this can only be applied to SN hosts, as all other
objects are deselected from the input target catalogue once a $Q=3$
quality redshift is obtained. There are 1100
SN hosts with a $Q=3$ quality redshift, potentially indicating an
error rate in addition to that noted above of 1-3\%.

The SN hosts also allow one to get a handle on the repeatability of
the redshift measurement. We first selected SN hosts that have at
least four redshift measurements with a quality flag of 3 or 4. We
then compute the RMS of the redshifts for each host. The median RMS is
0.00015. While this is indicative of the redshift uncertainty, it is
likely that it is a lower limit, as the measurements are not fully
independent: redshifts derived from data taken later include earlier
data. Indeed, the median RMS is a factor of two smaller than the
uncertainty reported in Yu15, who report 0.0003 for SN hosts using
redshifts that were obtained from data that were fully independent.

The impact of a redshift uncertainty of 0.0003 on the dark energy
equation of state parameter is negligible if the uncertainty is not
systematic. If the uncertainty is systematic, then the dark energy equation of state
parameter could be biased \citep{Calcino2017}.

\subsection{Redshift Outcomes}

The diversity of objects that were observed over the six years by
OzDES is large; from radio sources covering a broad range in magnitude
and redshift, to more restricted classes of objects, such as LRGs,
covering smaller magnitude and redshift ranges.

\begin{figure*}
 \includegraphics[width=8cm]{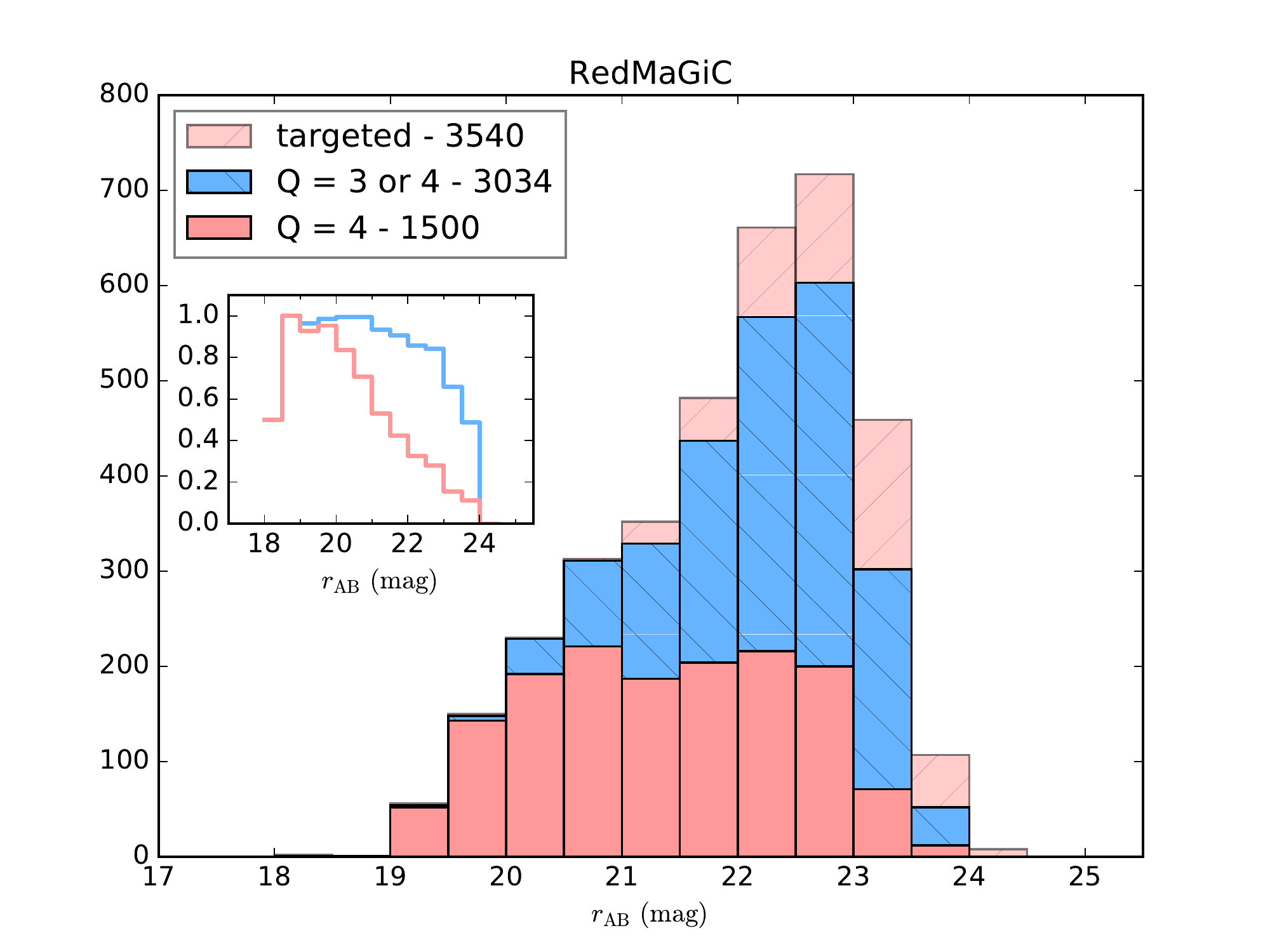}
 \includegraphics[width=8cm]{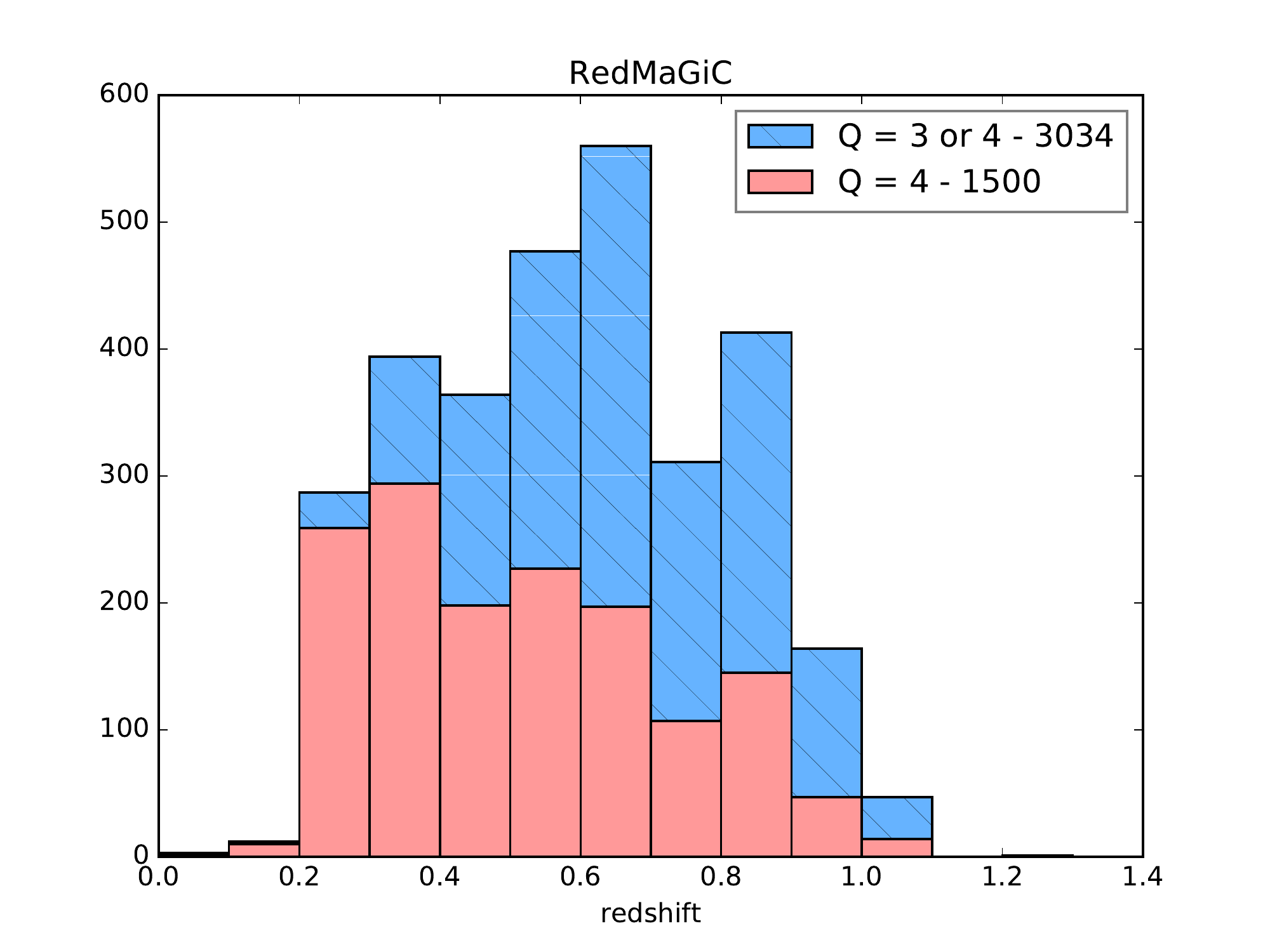}

 \caption{Left: The number of redshifts as a function of $r_{\mathrm
     AB}$ for redMaGiC galaxies. The inset shows the fractional
   completeness as a function of magnitude. The redshift completeness
   is the number of objects with Q=3 (or Q=4) divided by the number of
   objects that were targeted. For all object types other than host
   galaxies, sources were removed from the target input catalogues
   once they reached a Q=3 redshift quality. Right: Redshift
   histogram. RedMaGiC galaxies tend to be massive galaxies with
   little ongoing star formation, so their spectra are usually devoid
   of strong emission lines. Most of the redshifts were obtained from
   the calcium H and K lines in absorption, which are usually quite
   distinct in redMaGiC galaxies. }
 \label{fig:outomeRedMaGiC}
\end{figure*}

\begin{figure*}
 \includegraphics[width=8cm]{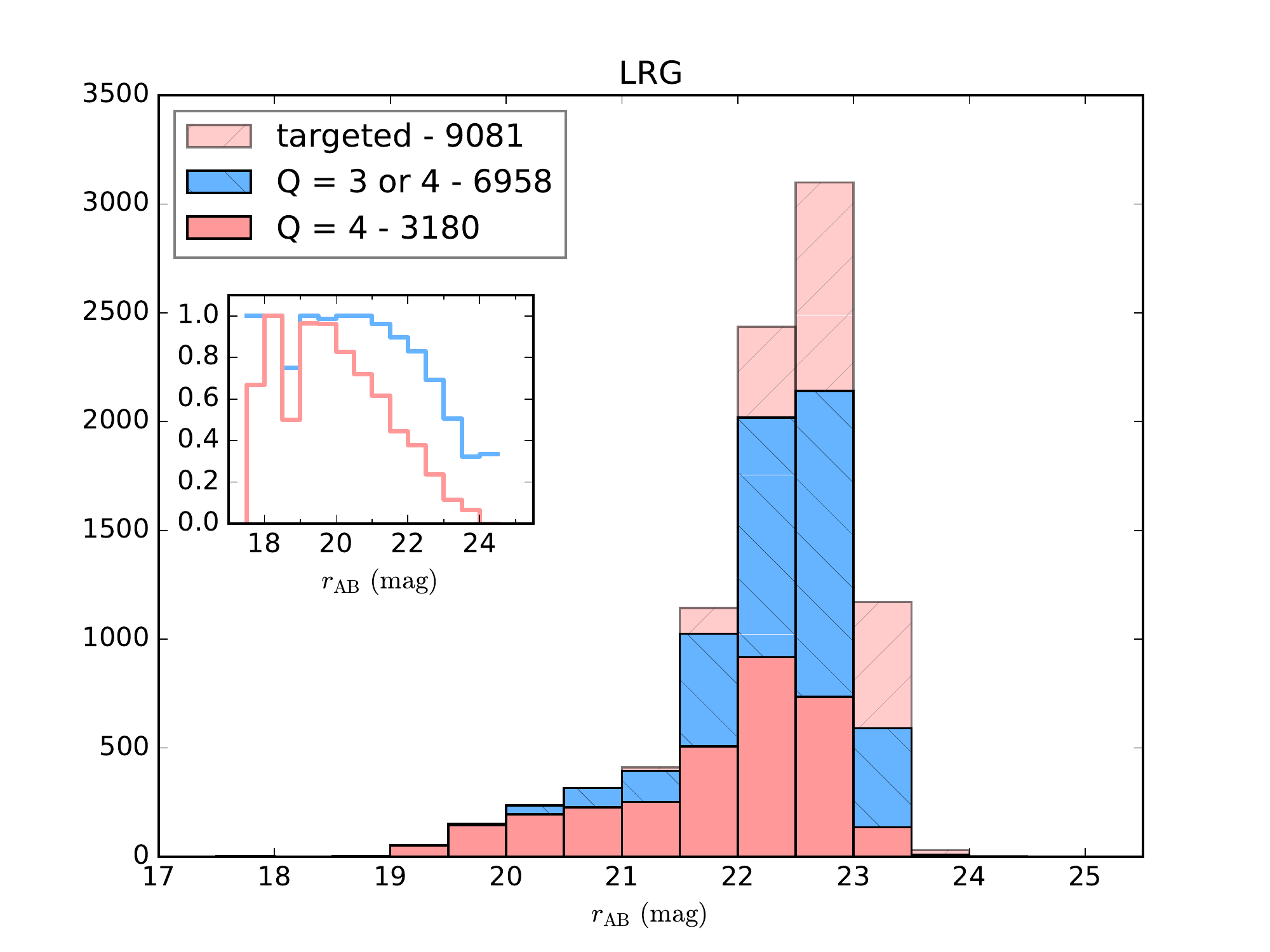}
 \includegraphics[width=8cm]{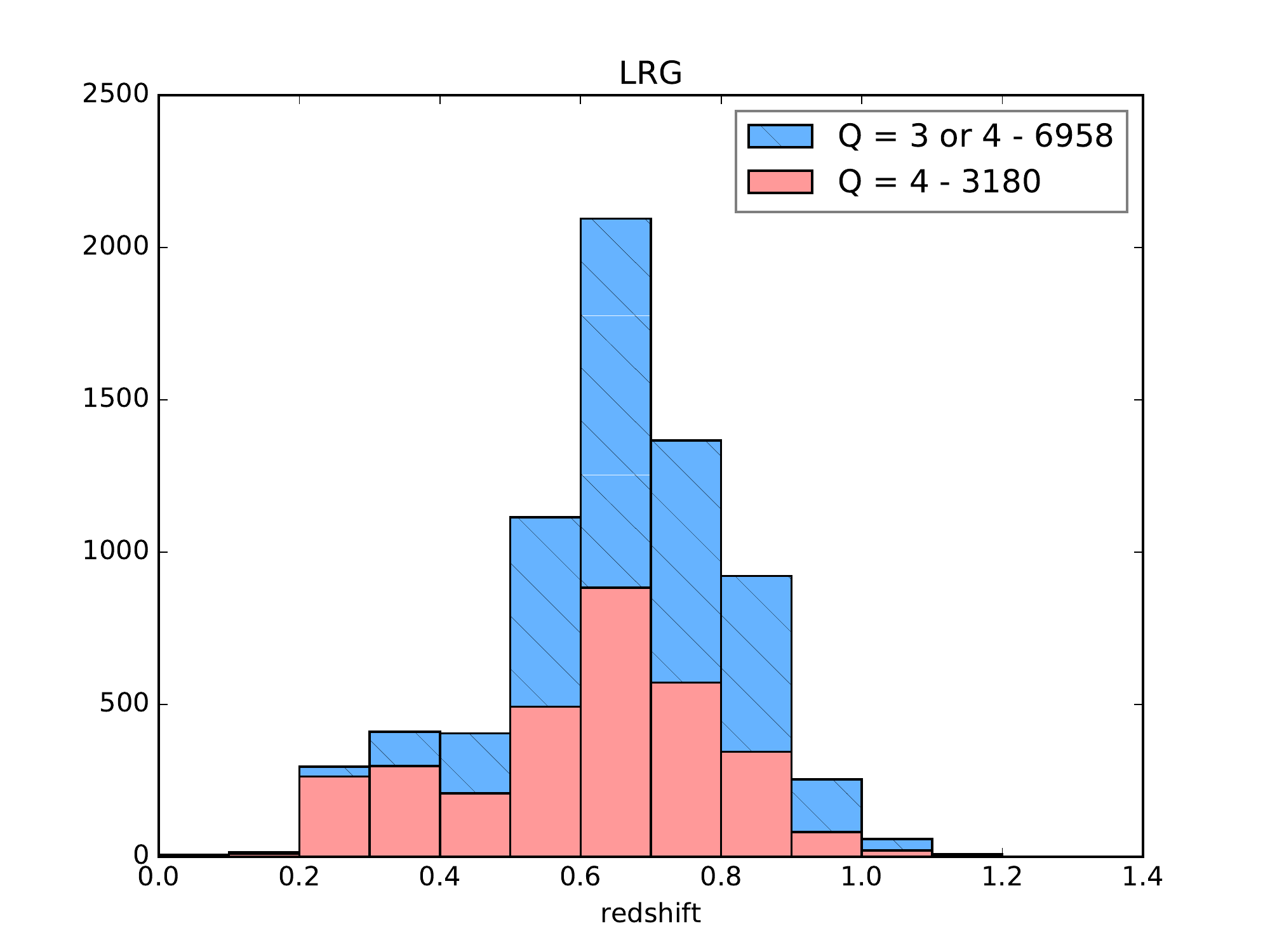}

 \caption{As for \protect{Fig.~\ref{fig:outomeRedMaGiC}} but for Luminous Red
   Galaxies. LRGs, like redMaGiC galaxies, also tend to have little
   ongoing star formation, so the spectra are usually devoid of strong
   emission lines.}
 \label{fig:outomeLRG}
\end{figure*}

\begin{figure*}
 \includegraphics[width=8cm]{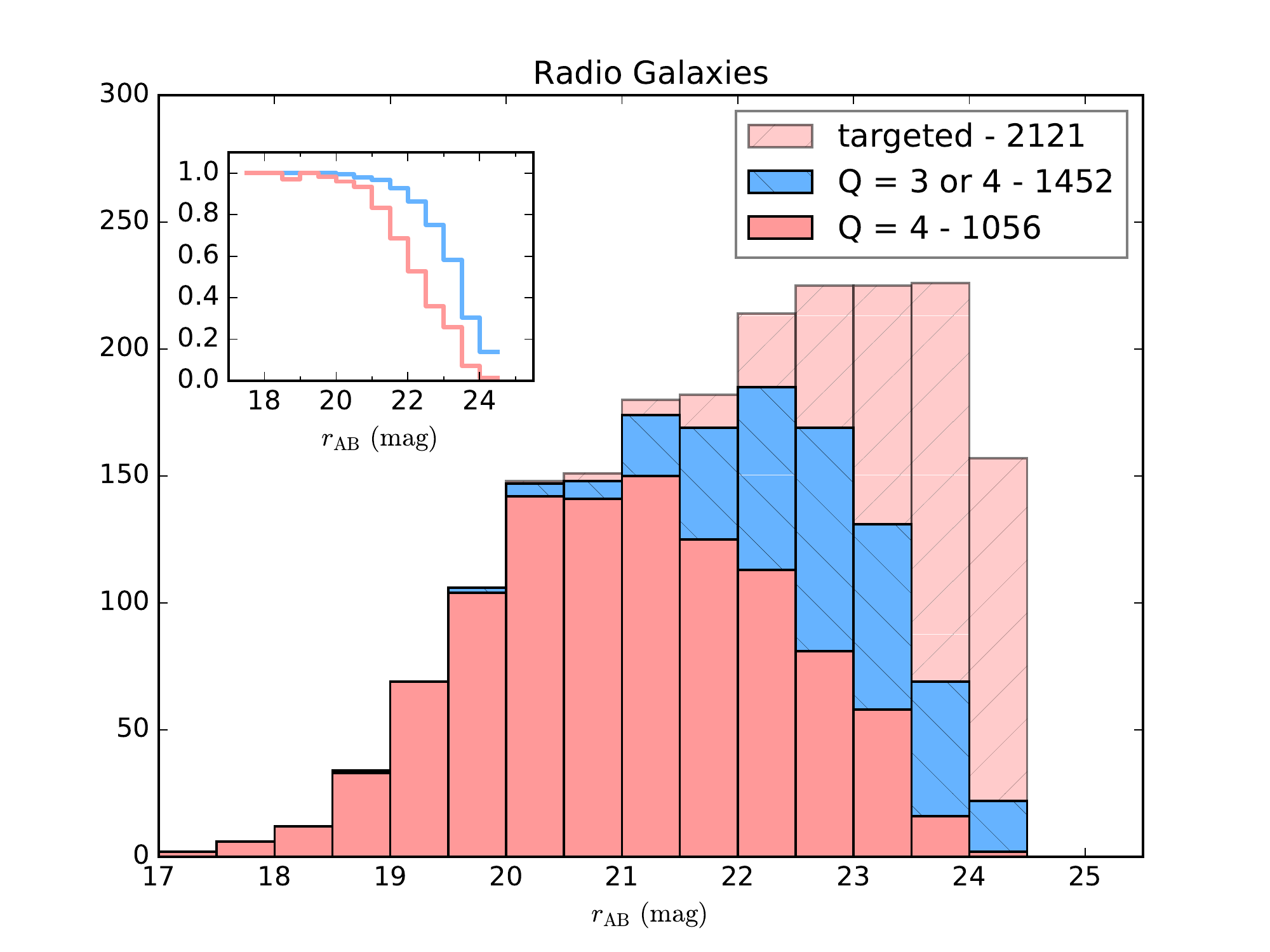}
 \includegraphics[width=8cm]{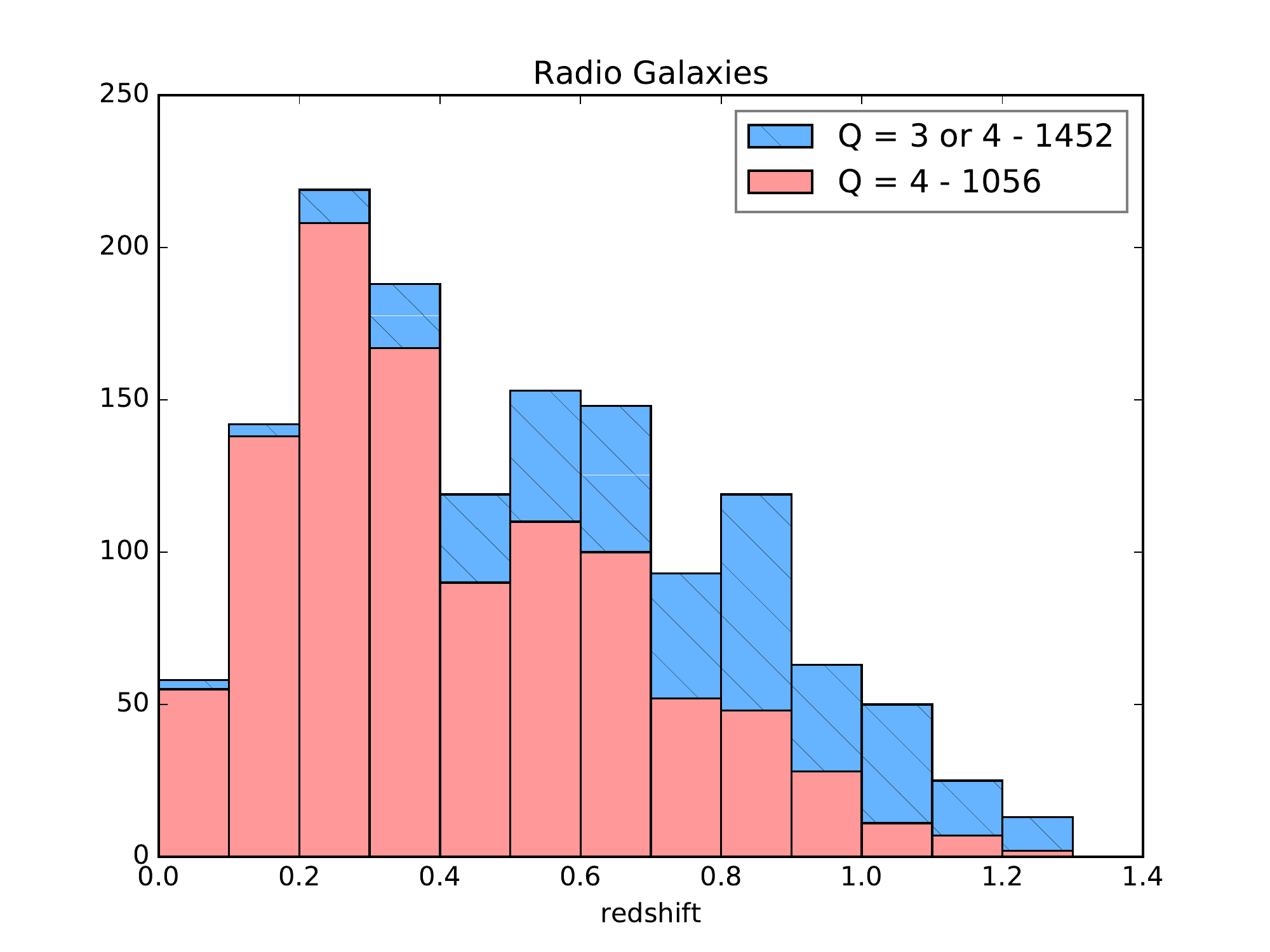}

 \caption{As for \protect{Fig.~\ref{fig:outomeRedMaGiC}} but for radio galaxies.}
 \label{fig:outcomeRadio}
\end{figure*}

In Figs~\ref{fig:outomeRedMaGiC}, \ref{fig:outomeLRG}, and
\ref{fig:outcomeRadio}, we provide magnitude and redshift
distributions for RedMaGiC galaxies, LRGs and radio galaxies.  As
noted at the end of Sec.~\ref{sec:allocation}, some targets had
multiple types. An LRG may have been classified as a redMaGiC galaxy
and visa-versa. About 40\% of all sources observed by OzDES had
multiple types. In these plots, we include all objects of the
indicated type, even if it also had another type.

Also shown as an insert to these figures and
Fig.~\ref{fig:completeness2} is the redshift completeness, defined as
the number of objects with a redshift divided by the number of objects
that were targeted. A number of targets were spectroscopically
classified as stars. These were removed before computing the redshift
completeness.

\subsection{Redshift Completeness}

\subsubsection{Completeness as a function of magnitude and exposure time}

A little more than one third of the 192,000 fibre hours that were
spent observing all targets were spent on SN hosts (see
Table~\ref{tab:allYears}). SN hosts were targeted during the entire
duration of OzDES and were not deselected from the input target
catalogue until a $Q=4$ quality redshift was obtained. They are
therefore useful in examining how redshift completeness depends on
magnitude and exposure time. However, caution is needed before
applying the results from SN hosts to the general population of field
galaxies, because SN hosts are a biased subset. SN hosts are more
likely to be $z<1.35$, and therefore within the spectral range covered
by OzDES.

All SN hosts were selected on the basis of the detection of a
transient in the DECam images. With the exception of super-luminous
supernovae, which are rare, there would be few SN hosts that are at
redshifts that places the [OII] $\lambda\lambda 3727$\AA\ doublet
beyond the red end of the spectral range covered by the setting we
used in AAOmega, which is at 8800\AA. The red end corresponds to a
redshift of $z\sim 1.35$ for [OII]. Nearly all of the SN Ia detected
by DES will be below this redshift \citep{Bernstein2012}. Hence, when
interpreting the completeness of OzDES as a function of magnitude, the
completeness will be different than for a survey that targeted field
galaxies.

\begin{figure}
 \includegraphics[width=\columnwidth]{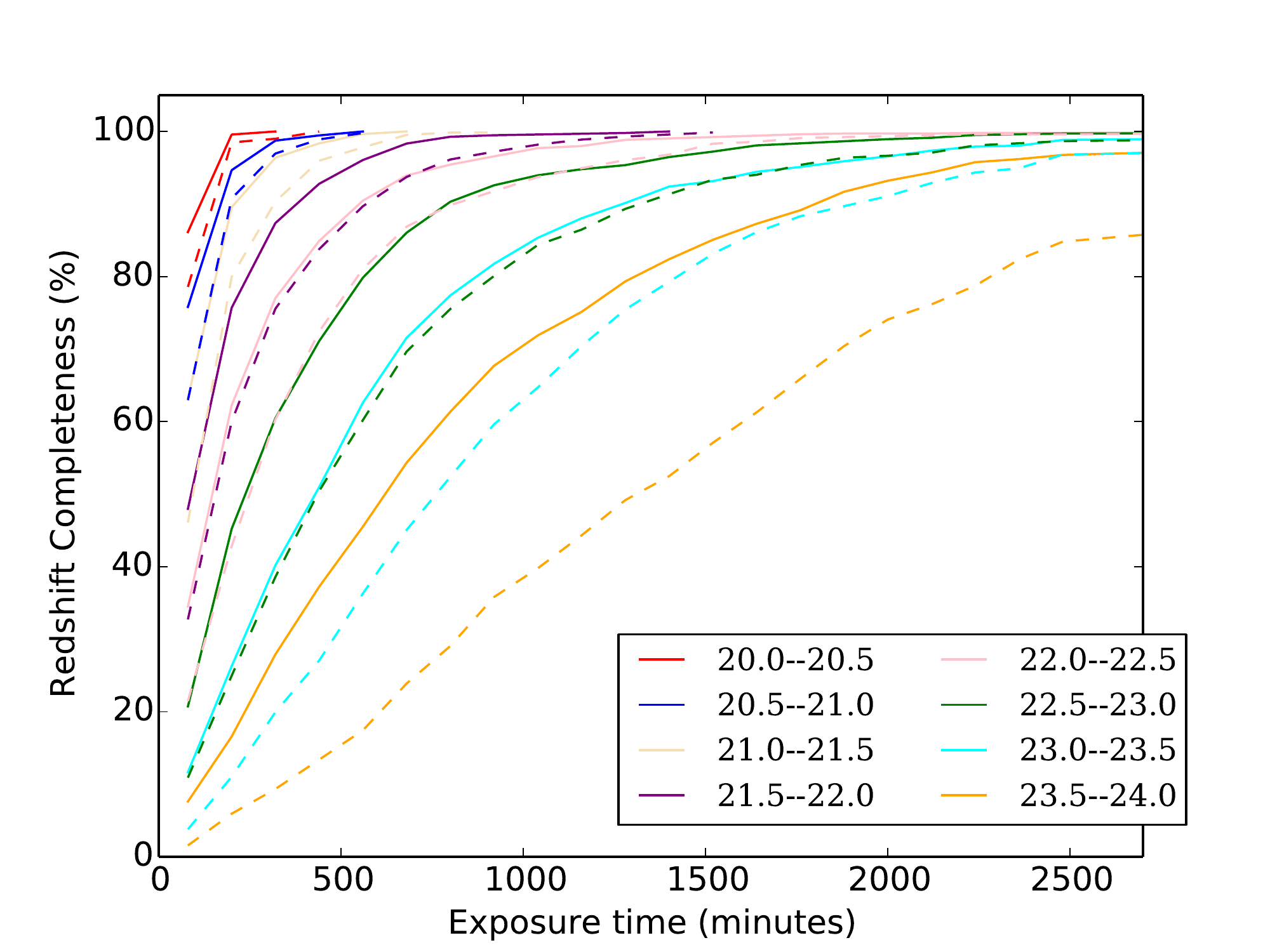}

 \caption{Redshift completeness as a function of exposure time for
   different magnitude bins and two redshift quality flags. Dashed
   lines represent redshift quality 4 and solid lines represent
   redshift quality 3. The lines end once completeness reaches 100\%.}
 \label{fig:completeness}
\end{figure}

The redshift completeness of SN hosts as a function of exposure time
is shown in Fig.~\ref{fig:completeness}. The completeness is shown for
different magnitude bins and two redshift quality flags. As expected,
completeness increases with exposure time and is highest for the
brightest sources. It is also higher for lower quality redshift flags.

\begin{figure}
 \includegraphics[width=\columnwidth]{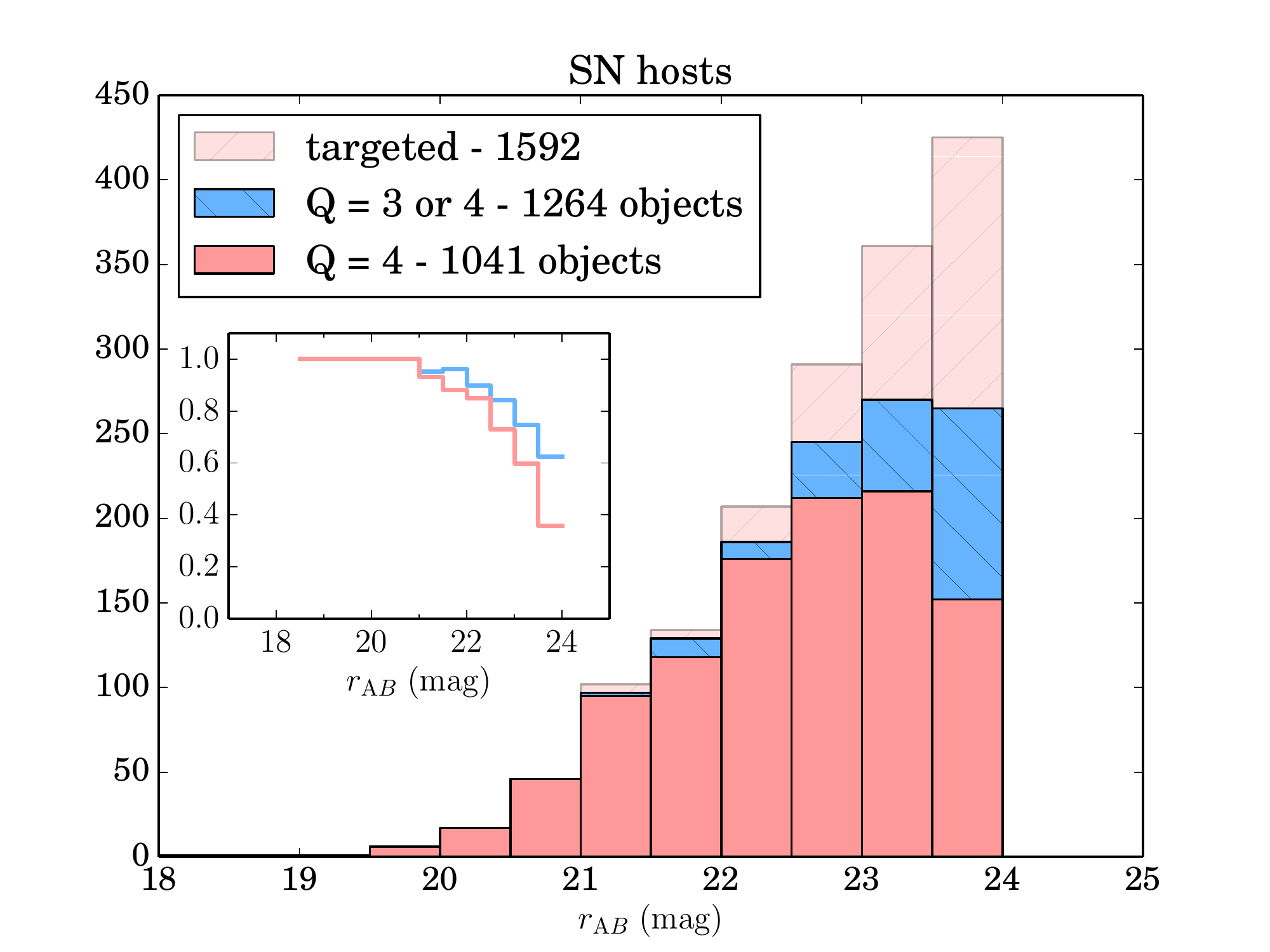}

 \caption{The number of SN hosts with redshifts binned by $r_{\mathrm
     AB}$. The redshift completeness is shown in the insert. Faint SN
   hosts (i.e. sources classified as {\tt SN\_host\_faint}) are not
   included. Throughout the survey, over 7,000 objects were targeted
   as potential SN hosts. Plotted here are the ones most likely to be
   SNe Ia. Excluded from this plot are 279 SN hosts that have
   redshifts from external surveys, as these objects were not
   targeted. The last bin would reach 95\% completeness if we were
   able to integrate for 40 hours on all sources, as shown in
   \protect{Fig.~\ref{fig:completeness}}.}
 \label{fig:completeness2}
\end{figure}

For the faintest bin ($23.5 < r_{\mathrm AB} < 24$), the redshift
completeness reaches $\sim95$\% for quality 3 redshifts after 40 hours
of integration. This does not mean that we reached this level of
completeness in the survey, as not all SN hosts have been observed for
that long. We anticipate that if we were to observe this long for all
sources, then we would reach this level of completeness. As shown in
Fig~\ref{fig:completeness2} the completeness in the faintest bin is
65\%.


\subsubsection{Completeness as a function of plate position}

The positioning of fibres relies on a series of transformations
between sky coordinates and plate coordinates. The transformations
include the {\tt poscheck}, which is updated by observatory staff,
typically during the first night after 2dF is remounted on the
telescope, and the {\tt plate survey}, which is run every time a field
is configured. The accuracy of these transformations has a direct
impact on the amount of flux entering fibres and on the redshift
completeness.

We examine how the redshift completeness varies across the field for
SN hosts and LRGs. For LRGs, we include redMaGiC galaxies. Any
variation in the redshift completeness across the field would be
indicative of a problem in the accuracy of the transformations.

\begin{figure*}
 \includegraphics[width=16cm]{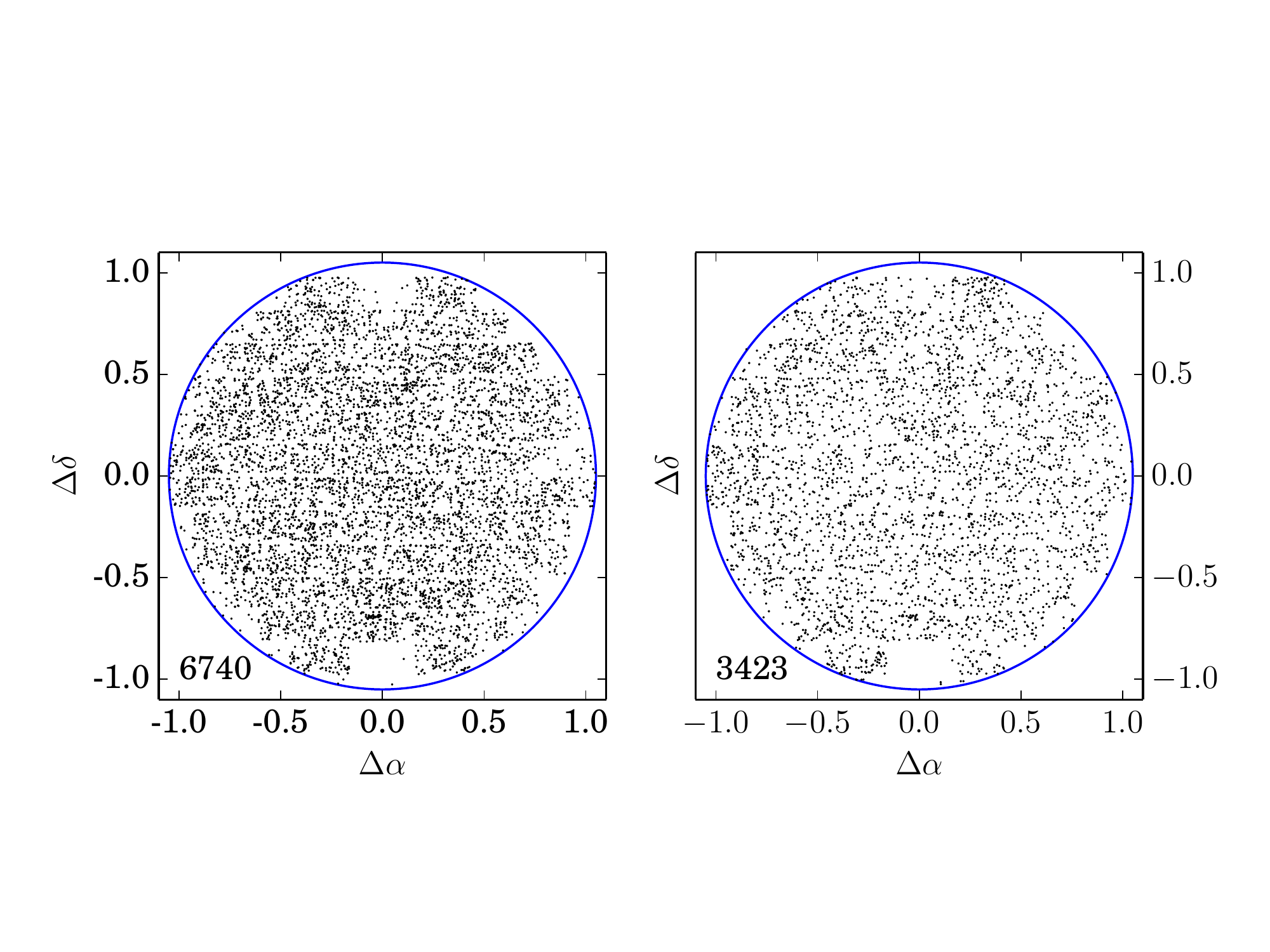}

\caption{The position, with respect to the centre of the 2dF field
  plates, of SN hosts with redshifts (left plot) and without redshifts
  (right plot). The blue circle represents the 2.1 degree diameter of
  the 2dF field of view. Notable in these images are the chip gaps,
  the boundary of the mosaic and the two CCD chips at the bottom and
  top of the mosaic that were not operational for some part of DES. In
  this plot North is up and East is to the right. Object numbers are
  given in the lower left corners.}
 \label{fig:SNcompleteness}
\end{figure*}

In the two panels of Fig.~\ref{fig:SNcompleteness}, we plot the
position of SN hosts with and without redshifts, respectively. One can
clearly see the boundary of the DECam chips traced by the SN
hosts. The 10 fields in the DES deep survey were observed with minimal
dithering. Consequently, chips gaps between the DECam CCDs were not
covered and so no SNe are found in these gaps. A deficit of SN hosts
can be seen in areas where CCDs were not functioning for some fraction
of the survey.

\begin{figure*}
 \includegraphics[width=16cm]{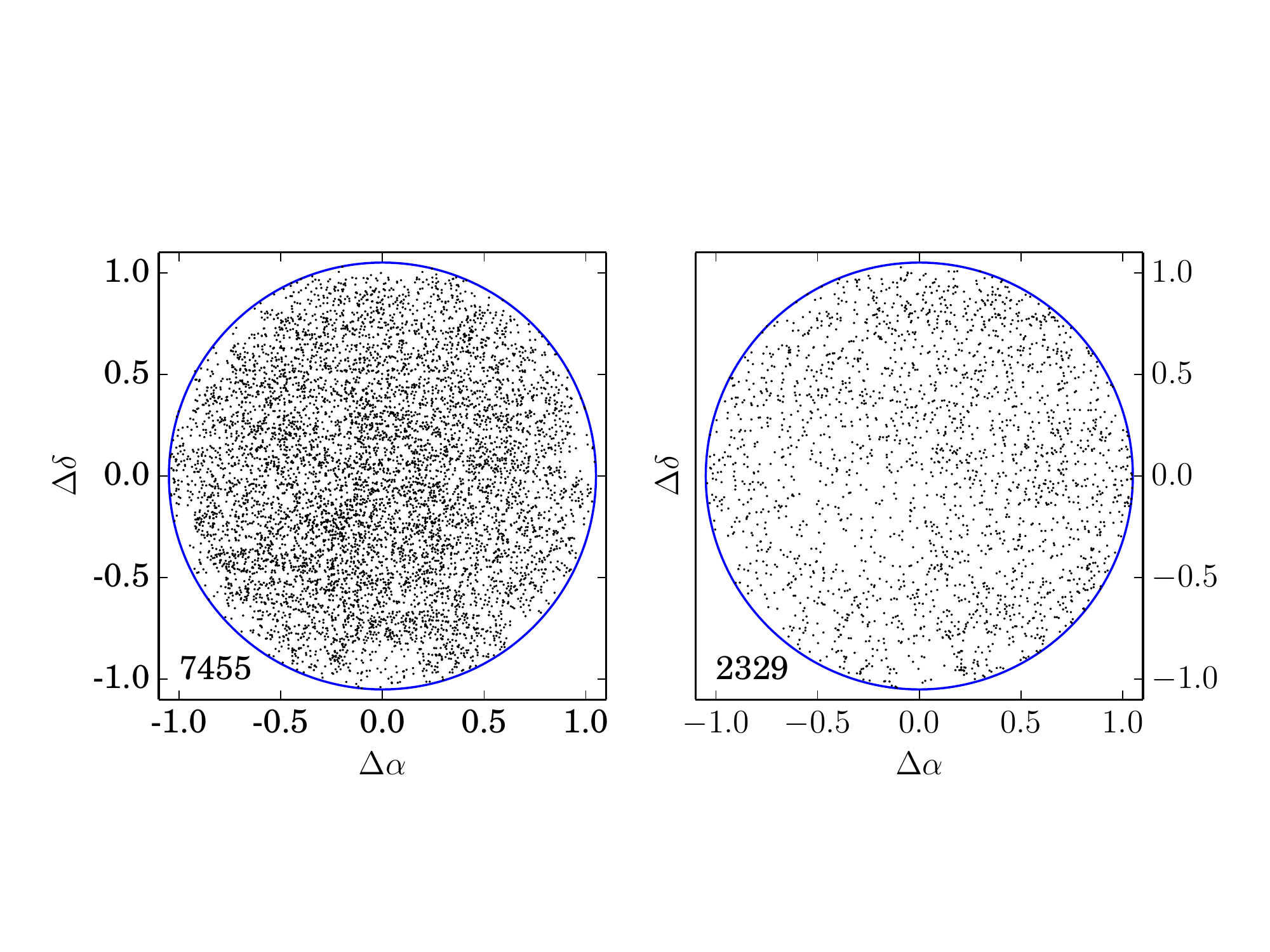}

 \caption{The position, with respect to the centre of the 2dF field
   plates, of LRGs with redshifts (left plot) and without redshifts
   (right plot). Gone are the chip gaps. The gap left by the chip at
   the bottom of the mosaic is still visible. Unlike the plot for SN
   hosts, there appears to be a gradient, which becomes clearer in
   \protect{Fig.~\ref{fig:Completeness}}. Object numbers are given in the lower
   left corners.}
 \label{fig:LRGcompleteness}
\end{figure*}

Similarly, in the two panels of Fig.~\ref{fig:LRGcompleteness}, we
plot the position of LRGs with and without redshifts. The boundary of
the DECam chips can no longer be seen, as LRGs were selected from
imaging data that was more continuous.

\begin{figure*}
 \includegraphics[width=8cm]{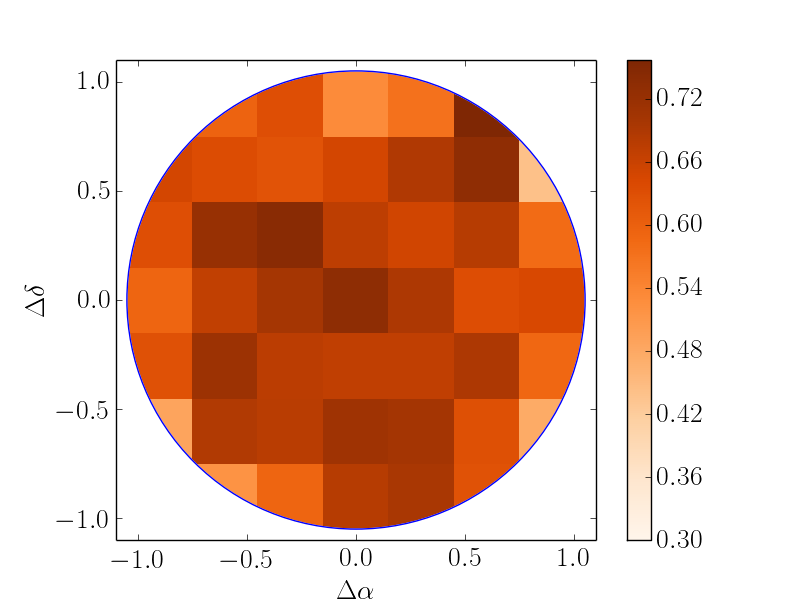}
 \includegraphics[width=8cm]{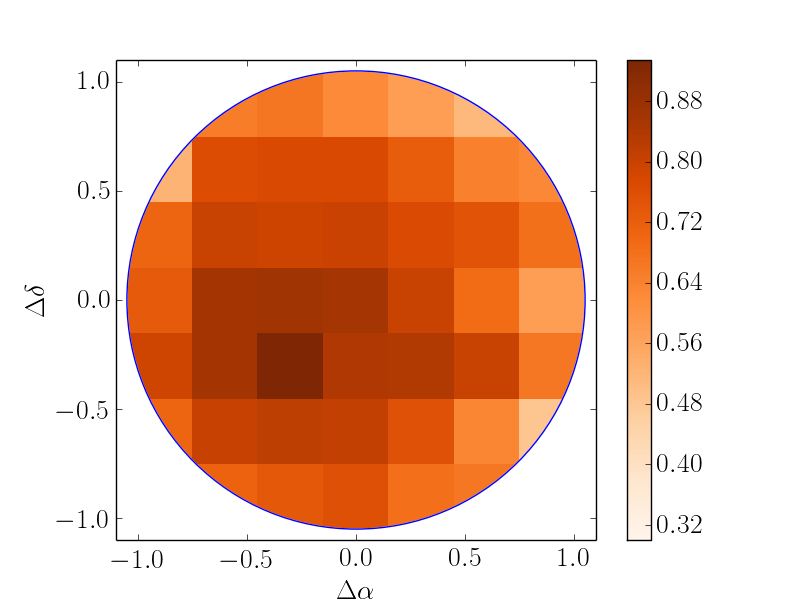}

 \caption{Redshift completeness in bins of 0.3 degrees by 0.3 degrees
   for SN hosts (left) and LRGs (right). The blue circle represents
   the 2.1 degree diameter of the 2dF field of view. The decreasing
   redshift completeness with distance from the centre of 2dF is a
   consequence of poorer image quality at larger radii. The presence
   of a gradient in the right plot and not the left one is
   surprising.}
 \label{fig:Completeness}
\end{figure*}

The redshift completeness is shown for SN hosts and LRGs in
Fig.~\ref{fig:Completeness}. The completeness is not uniform for
either class of object. In both cases, there is a drop towards the
edge of the field. This is not unexpected and is due a drop in the
image quality delivered by the 2dF corrector at the edge of the field
and to the accuracy at which the plate can be rotated, which affects
fibre near the field edge more than fibres in the centre. Less
expected is a gradient from lower left to upper right in the redshift
completeness of LRGs

\begin{figure*}
 \includegraphics[width=8cm]{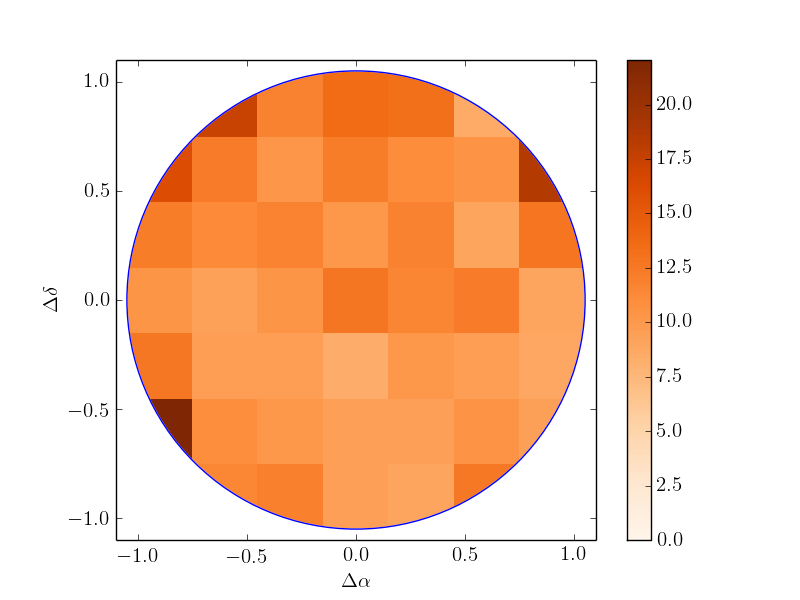}
 \includegraphics[width=8cm]{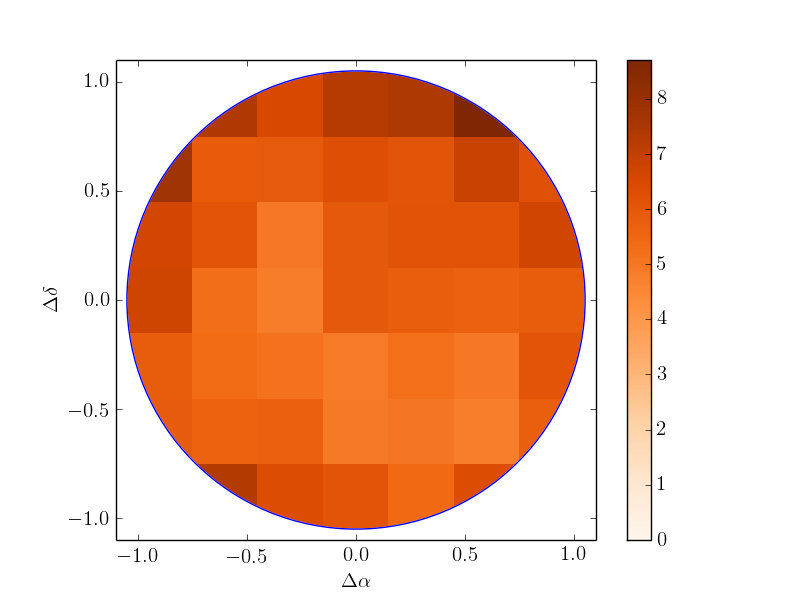}

 \caption{The average number of exposures, in bins of 0.3 degrees by
   0.3 degrees, required for a successful redshift measurement for SN hosts (left) and LRGs (right). The blue circle
   represents the 2.1 degree diameter of the 2dF field of
   view. Overall, as one approaches edges of the 2dF field of view, it
   takes more exposures to secure a redshift. The gradient from lower
   left to upper right seen in \protect{Fig.~\ref{fig:Completeness}} is also
   seen here and is also stronger for the LRGs.}
 \label{fig:AverageExposureTime}
\end{figure*}

In Fig~\ref{fig:AverageExposureTime}, we show the average number of
exposures it takes to get a redshift. There is a general trend of
increasing exposure time as one moves from the centre of the
field-of-view to the edges. Again, the LRGs show a more significant
gradient than the SN hosts.

There are a couple of potential reasons why the gradient is less
significant for SN hosts. Firstly, SN hosts are not removed from the
target catalogue until a secure, quality 4 redshift is obtained,
whereas LRGs are removed from the target catalog once a quality 3
redshift is obtained. This will reduce field-dependent redshift
incompleteness. Secondly, the flux lost due to fibre positioning
errors is more severe for point sources than extended objects. LRGs
due to their higher redshift and redder colours, will be more point
like than SN hosts. At the AAT the profile of sources at these
redshifts is dominated by the seeing, which will be better in the red
than the blue.

The gradient has no impact on the primary OzDES science goals, as the
spatial location of the AGN and SNe are not relevant, but the gradient
needs to be taken into account if these data were to be used to
estimate clustering from statistics such as the 2-point galaxy-galaxy
correlation function.

\section{The second OzDES Data Release}\label{sec:datarelease}

This paper marks the second OzDES release, which will be referred to
as OzDES-DR2. OzDES-DR2 differs from OzDES-DR1\footnote{The redshift catalogue for OzDES-DR1 is available
  from \url{https://datacentral.org.au}.}, the first OzDES data release,
in two key aspects.

\begin{itemize}
    \item The redshift catalogue includes all objects observed during
      the 6 years of OzDES and the OzDES pre-survey. This includes SN
      hosts, SNe, and targets for which a redshift could not be obtained.
    \item The fully processed spectra are provided, including all individual spectra, as well as the stacked spectrum of each object. 
\end{itemize}

In OzDES-DR2, we release 375,000 spectra of almost 39,000 objects, of
which three quarters have a redshift with a quality flag of 3 or
greater. We release all spectra, including the spectra of objects that
do not have a redshift. These data can used to help secure redshifts
for these objects if they are observed in the future with the AAT or
other spectroscopic facilities. A full description of the data release
is provided in Appendix~\ref{sec:DR}.


\section{AGN reverberation mapping}\label{sec:AGNRM}

Second to SN hosts, AGN were the next most frequently observed target
type. As shown in Fig.~\ref{fig:fullallocations}, about a quarter of
the fibre hours available to OzDES were allocated to the AGN
reverberation mapping (RM) programme. The first year of the OzDES RM
programme was used to define the sample that was followed in the
subsequent five years. Details of the selection are provided in
\citet{King2015}. In brief, AGN were selected on the basis of the
quality of the spectra that were taken in Y1, on the presence of
emission lines, while avoiding AGN that occasionally fell in chip
gaps, maximising the number of AGN with two reverberating spectral
lines in our wavelength range, and ensuring that the AGN selected
sampled the redshift interval $0 < z < 4$. Most AGN that pass these
cuts have have $r_{\mathrm AB} < 21.2$. From an initial sample of 3331
AGN, 771 AGN were followed during the next five years.

Apart from active transients, AGN in the OzDES RM sample have the
highest priority, so the percentage of AGN that were allocated a fibre
in any given observation was near to 100\%. However, not all fields
could be observed during every OzDES run, because of a combination of
the finite duration of the observing runs, time lost to poor observing
conditions, and the differing priority of the ten DES deep
fields. Considering the latter and the bar chart shown in
Fig.~\ref{fig:frequency}, we expect that the AGN in the two E-region
fields and the X3 and C3 fields will have the greatest number of
observations, both in terms of the raw number of times they were
observed, and the number of runs in which they were observed.

\begin{figure}
 \includegraphics[width=8cm]{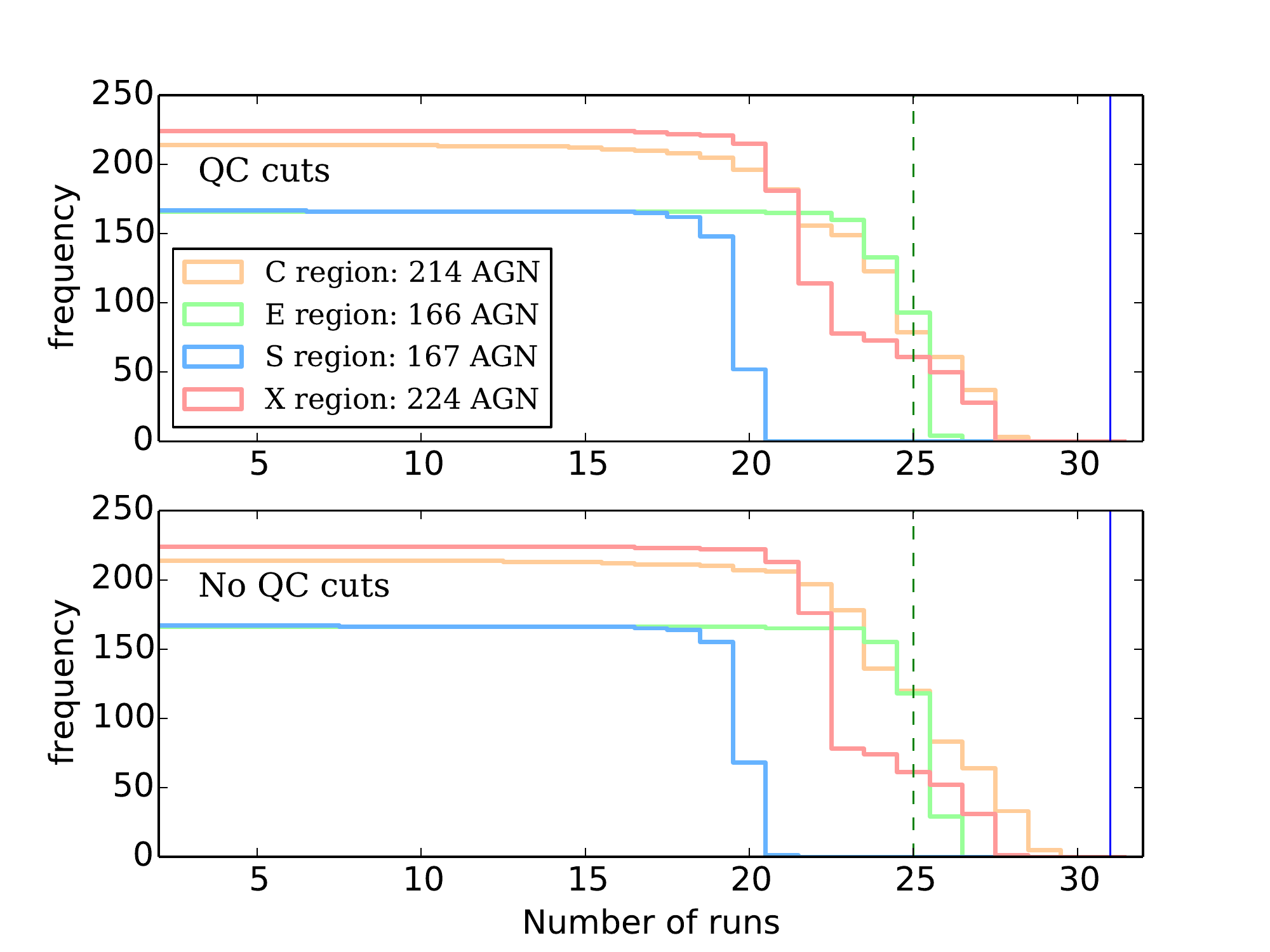}

 \caption{Reverse cumulative histograms of the number of independent
   spectroscopic epochs in the AGN reverberation mapping programme,
   plotted without restriction on data quality (lower histogram) and
   excluding poor quality data (upper histogram). See text for
   details. The vertical line corresponds to the number of runs on
   which some OzDES data were taken during the six years OzDES ran. It
   corresponds to the maximum possible number of epochs. The
   prediction on the number of recovered lags made in \protect{\citet{King2015}}
   were based on obtaining 25 epochs, which is shown as the dotted
   line.}
 \label{fig:AGNepochs}
\end{figure}

In Fig.~\ref{fig:AGNepochs}, we show the distribution of the number of
runs in which AGN were observed. Two plots are shown. In the lower
plot, all observations are included, regardless of the observing
conditions or the quality of the data. In the upper plot, only data
that had a relative zero point greater than $-2$ (see
Fig.~\ref{fig:throughput}) and a quality control (QC) flag of `ok' are
included (see Appendix B for a discussion on the QC flag). The
vertical solid line shows the number of runs on which OzDES
observations were scheduled during the six years that OzDES ran. The
two runs from the OzDES pre-survey are excluded, since the AGN RM
pre-sample had not been defined at that time. This solid line
represents an upper limit to the number of epochs. The median values
for the four regions are shown in Table~\ref{tab:AGNepochs}. The
median values for the X and C regions include X3 and C3, which were
the highest priority fields, so we list the median values for the C3
and X3 separately as well. For the E region fields and X3 and C3, the
median is close to the 25 epochs that were used in the simulations of
the OzDES RM programme in \citet{King2015}.

With this number of epochs, simulations in \citet{King2015} predict
that one should recover lags for 35-45\% of the AGN depending on the
accuracy with which one can calibrate the line flux. The simulations
used a 10\% accuracy for the line measurements.  \cite{Hoormann2019}
demonstrate that we are able to achieve this level of accuracy, so we
anticipate that we will recover this fraction of lags in the two E
fields and the C3 and X3 fields, and a lower fraction in the other six
fields. The first two lags, using the CIV line and based on a subset
of the OzDES data, were published in \cite{Hoormann2019}.

\begin{table}
\begin{center}
 \caption{Median number of AGN epochs.}
 \label{tab:AGNepochs}
 \begin{tabular}{ll}
  \hline
   Fields & Epochs\\
    \hline
C1, C2, and C3 & 23 \\
E1 and E2 & 24 \\
S1 and S2 & 18 \\
X1, X2, and X3 & 21 \\
C3 & 25 \\
X3 & 25\\
  \hline
   \hline
 \end{tabular}
 \end{center}
\end{table}

\section{SN host galaxies: A case study}\label{sec:SNhosts}

As illustrated in Fig.~\ref{fig:fullallocations}, slightly more than
one-third of all the fibre hours allocated to targets during the six
years of OzDES were allocated to SN hosts. Unlike other targets, SN
hosts were observed until a secure redshift (see
Sec.~\ref{sec:redshifts}) was obtained. SN hosts also extend to
fainter magnitudes than most targets. Both of these factors meant that
the integration time for some hosts are very long, as long as 2 days
in a number of cases.

We use SN hosts to examine how well the measured noise changes with
time, and explore the implications of deselecting SN hosts once they
reach a $Q=3$ quality redshift and before they reach the more secure
$Q=4$ quality redshift.

\subsection{Dependence on the measured noise with integration time}

Ch17 showed that the noise in the spectra evolves with increasing
exposure time, $t$, according to expectations. The OzDES pipeline
averages spectra from individual exposures rather than summing them,
so the measured noise decrease as $\sqrt{t}$, or with the square root
of the number of exposures if the exposure time is constant. Nearly
all OzDES exposures were taken with an integration time of 40
minutes. We repeat the Ch17 analysis, which was done with just three
years of OzDES data. With six years of data, there are more objects in
total and there are more objects with longer integration times.

We do apply some qualitative cuts to the data chosen for this
analysis. We exclude spectra that i) were taken in poor conditions
(defined here as exposures that have a relative zero point of less
than $-2$; see Fig~\ref{fig:throughput}), ii) did not pass quality
control, and iii) were taken during Y1 and SV. During Y1 and SV, we do
not have reliable zero points, and did not take dome flats, so the
quality of the processed data is not as good as it was in later years.

With these cuts, we then select SN hosts that have at least 12
exposures. This will tend to select SN hosts that are fainter than
average, as redshifts for bright hosts would have typically been
obtained with fewer exposures and would have been deselected from the
target pool before 12 exposures had been taken. As the selected
objects are faint, the signal to noise ratio is quite small, so
spectral features do not contribute to the noise.

We then select a wavelength region over which to measure the noise. We
start off with a region that is relatively free of night sky
lines. The region we select starts at 6610\AA\ and ends at 6750\AA. We
then examine how our results change if we choose a region that
contains bright night sky lines.

We then randomise the order in which exposures were taken. The
ordering is different for each object. For a given object, we take the
first exposure and compute the clipped RMS over the wavelength region
noted above. In computing the clipped RMS, we reject the 5 highest and
5 lowest values. Over the wavelength region that we sample, there are
about 135 pixels, so the clipping removes less than 10\% of the
data. The clipping allows us to exclude residuals from cosmic rays
that were not perfectly removed during the processing of the data and
mitigates the effect of strong spectral features, such bright emission
lines from the object. We experimented with the normalised median
absolute deviation instead of the clipped RMS and obtained
qualitatively similar results. All subsequent measurements of the
clipped RMS are scaled by the clipped RMS that is computed in this
first exposure.

\begin{figure}
 \includegraphics[width=8cm]{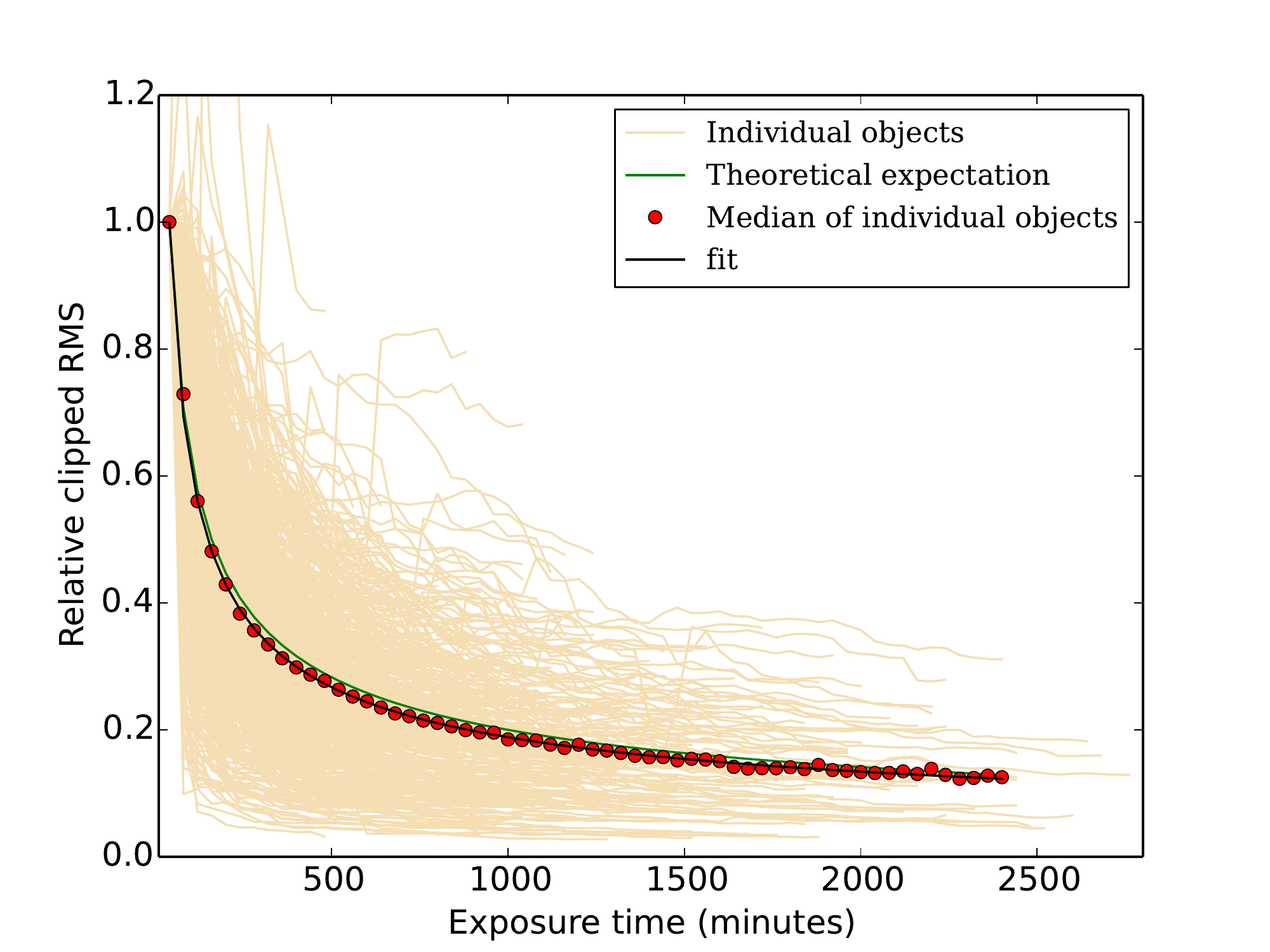}

 \caption{The scaled clipped RMS of 639 individual objects (in light
   yellow) from the three C fields plotted against exposure time. The
   median of these curves is plotted as the solid red circles. The
   theoretical expectation is the green curve and the best fit is the
   black curve. Note how closely the two match. The clipped RMS is
   computed over the region 6610-6750\,\AA.}
 \label{fig:SNratio}
\end{figure}

We then add data from subsequent exposures in the randomised sequence,
noting that we average the spectra, rather than summing them, and
recompute the clipped RMS. This is done for every SN host. The results
are shown in Fig~\ref{fig:SNratio}.

There is considerable scatter from object to object. This is caused by
data of poorer than average quality entering the average at different
stages of the sequence. For example, a sequence of exposures in which
the first exposure is noisy will result in a much larger than
anticipated drop in the relative clipped RMS when data of better
quality is included. Applying stronger quality cuts reduces the
scatter. To get an idea of the general trend, we take the median of
all objects. These are the red circles in Fig.~\ref{fig:SNratio}.

We fit the function

\begin{equation}  \frac{n^{\alpha}+\beta}{1+\beta}\label{eq:SNratio} \end{equation}

to the red points. The function equals 1 when $n=1$, and allows for a
noise floor. The theoretical expectation for Poisson noise is
$\alpha=-0.5$ and $\beta=0.0$. This is represented with the green
curve. The best fit to Eq.~\ref{eq:SNratio} is the black curve. The
best fit has $\alpha=-0.54\pm0.01$ and $\beta=0.02\pm 0.01$. The best
fit is close to theoretical expectation for Poisson noise, and the
floor is low. We find there is no reason to stop observing targets
over the range of exposure times explored here. We will explore this
point further in the next section.

There are limitations to our approach. Eq.~\ref{eq:SNratio} may not be
the best functional form, and we have not allowed for the hard limits
imposed by theory (i.e $\beta \ge 0$ and $\alpha \ge -0.5$) when
fitting Eq.~\ref{eq:SNratio}. Indeed, we find $\alpha < -0.5$. It is
unclear what causes this.

We varied our selection cuts and repeated the analysis for the other
fields to see how robust our results were. We do see some variability
in the results, the origins of which are unclear. For the permutations
we tried, we found $-0.55 < \alpha < -0.5$ and $0 < \beta < 0.05$, i.e. a
slightly steeper dependence than expected from theory and weak
evidence for a noise floor. The floor is still less that the
contribution from Poisson noise even after a day of exposure. At 1.5
days (which corresponds to 54 40-minute exposures), $54^{-0.5} \sim
0.14$.

We also varied the wavelength region over which the statistics were
computed, using both a region free of bright night sky lines
(6610-6750\,\AA) and a region dominated by them (7700-8000\,\AA). In
both cases, a dependence close to $1/\sqrt{n}$ was observed.

\subsection{Deselecting targets}

It is clear that, for the range of exposure times explored in OzDES
(up to two days of integration), the signal-to-noise continues to
improve as expected, i.e. with the square root of the exposure time.
If OzDES were to continue into the future, then exposure times could
grow into many days. A number of questions then arise. How long should
one spend on a target before deselecting it for a another one if one
is unable to get a redshift, and is there a way of knowing if some
targets are more likely to result in a redshift than others? Any
decision to deselect one source in preference for another needs
careful consideration, as such a decision will invariably lead to
biases that need to be modelled.

During Y6, when 71\% of the available fibre time was spent targeting
SN hosts, there were more SN hosts without quality 4 redshifts than
2dF fibres, especially for C3 and X3, the two deepest DES fields. If
there were many more fibres than the 392 available with 2dF, and SN
hosts were the only targets, then the answer would be to continue
observing these targets.

For instruments like 4MOST, which has 4 times as many fibres as 2dF,
one could observe more targets longer without running out of
fibres. Of course, 4MOST could choose to push to fainter limits than
OzDES did (most OzDES hosts are brighter than $r_{\mathrm AB}=24$), in
which case it too would have to make a choice about what to keep and
what to drop in favour of something that has a higher chance of
success.

Ultimately, this decision has to be based on the questions one is
trying to address. A SN with a noisy light curve at $z\sim 1$ may not
be as valuable as two SNe with high quality light curves at a redshift
of a half. We do not address this question in this paper, but leave it
to when the DES 5-year photometrically selected sample is defined.

For SN hosts, there is an additional consideration. SN hosts are
deselected once they get a $Q=4$ quality redshift. Some SN hosts
obtain a $Q=3$ quality redshift relatively quickly, but fail to reach
the $Q=4$ quality redshift that is necessary for deselection. This
raises an interesting question. How long should one continue observing
such targets once the $Q=3$ quality redshift is obtained?

We examine the number of exposures SN hosts remain at $Q=3$ once that
quality is obtained. We divide these objects into two. Those that
eventually obtain a $Q=4$ quality redshift and those that do not.
Almost 85\% of objects that move from $Q=3$ to $Q=4$ do so within 10
exposures of obtaining $Q=3$. If one were to deselect targets that
fail to reach $Q=4$ within 10 exposures of obtaining $Q=3$, then one
would free up 6.6\% of the fibre hours, which could then be used on
other sources, at the cost of obtaining 6.2\% fewer $Q=4$ quality
redshifts.

In OzDES, we did not deselect SN hosts before they reached $Q=4$, no
matter how many exposures were being taken. We are now in position to
explore how the choice of deselecting SN hosts biases the fitted
cosmological parameters. This will be reported in a future paper.

About 35 SN hosts fainter the $r=23$ fail to obtain a $Q=4$ quality
redshift once a $Q=3$ quality redshift was obtained, even after
continuing to integrate for 30 or more exposures (equivalent to 20
hours of integration). We examined each of these sources, and found
that more than half of them show a single bright line, which is
assumed to be the [OII] doublet. These objects reach the $Q=3$ level
relatively quickly. The redshift of these sources is around one,
meaning that other potential emission lines were outside the spectral
range of AAOmega. During run041, we used a redder setting for AAOmega,
which allowed us to pick up the [OIII] doublet in some cases. Most of
the remaining cases consist of sources with weak features, typically a
combination weak [OII], [OIII], and $H_{\beta}$, that only start to
become apparent after many hours of integration, but are not yet
strong enough for a $Q=4$ quality redshift to be assigned, even after
an additional 20 hours of additional integration.

\section{Implications for future surveys}

By the early 2020s, LSST will have started surveying the Southern sky
and several new multi-object spectroscopic facilities fed by rapidly
configurable fibre positioners covering wide fields of view will be
operational. Of the planned surveys using these new facilities, the
Time-Domain Extragalactic Survey \citep[TiDES,][]{Swann2019}, using
4MOST \citep{deJong2019} in combination with LSST, is the one that
most closely matches the OzDES concept. A comparison of the OzDES and
TiDES surveys is shown in Table~\ref{tab:TiDES}. The survey
characteristics of TiDES listed in Table~\ref{tab:TiDES} are subject
to change as the survey strategy is still evolving.

\begin{table}
 \caption{A comparison between TiDES/4MOST and OzDES/AAOmega.}
 \label{tab:TiDES}
 \begin{tabular}{lll}
  \hline
Metric & OzDES/AAT & TiDES/4MOST\\
\hline
\multicolumn{3}{c}{Telescope, Spectrograph, and Positioner}  \\
\hline
M1 Diameter (m) & 3.9 & 4.2\\
Field of view (sq. deg.)& 3.46 (circle) & 4.2 (hexagon) \\
Fibre diameter (\arcsec)& 2.0-2.1 & 1.45 \\
Number of fibres & 392 & 1624$^a$\\
Config. time (min) & 40$^b$ & 2\\
Spectral Resolution  & 1400-1700 & 6500\\
Wavelength range (\AA) & 3700-8800 & 3700-9500 \\
Smallest sep. (\arcsec)& 30 & 15 \\
\hline
\multicolumn{3}{c}{Survey characteristics}\\
\hline
Survey region$^{c,d}$ & DES deep fields & LSST DDF and  \\
& & $+5^{\circ} < \delta < -70^{\circ}$ \\
Median Seeing (\arcsec)& 1.5 & 0.8 \\
Usable clear fraction & 67\% & 90\% \\
Duration (years) & 5$^g$ & 5 \\
Transient redshifts & 5,000 & 50,000 \\
AGN monitored & 771 & 700 \\
Live transients & 1,450 & 30,000$^e$\\
Fibre hours & 193,000 & 250,000\\
Freq. (per lunation)  & 1 & 2$^f$ \\ 
Observing season & 5 months & 9 months$^f$ \\
\hline
\multicolumn{3}{l}{\footnotesize$^a$ Fibres feeding the two low-resolution spectrographs.}\\ 
\multicolumn{3}{l}{\footnotesize$^b$ Done in parallel while observing with the opposing plate.}\\ 
\multicolumn{3}{l}{\footnotesize$^c$ The 10 DES deep fields cover 27 sq. degrees.}\\
\multicolumn{3}{l}{\footnotesize$^d$ Deep Drilling Fields} \\
\multicolumn{3}{l}{\footnotesize$^e$ Mostly from the TiDES wide survey} \\
\multicolumn{3}{l}{\footnotesize$^f$ In the deep drilling fields.}\\
\multicolumn{3}{l}{\footnotesize$^g$ The SN programme in DES and OzDES lasted 5 seasons.} \\

 \end{tabular}
\end{table}

On practically every metric, TiDES will be superior to OzDES. It will
observe at least 10 times more live transients, mostly in the wide
survey, it will measure up to 10 times more host galaxy redshifts, and
it will monitor AGN at twice the frequency that OzDES did and with
shorter seasonal gaps. TiDES will also observe transients over the
majority of the southern sky, something that OzDES did not do.

Although the science goals and strategy of TiDES are heavily based
upon that of OzDES, practically there are some differences. The first
difference between OzDES and TiDES is that TiDES will not drive the
pointings of 4MOST. Wherever 4MOST is scheduled to point there will be
previously discovered transient targets, a majority of these targets
anticipated to be discovered by LSST. If these transients are still
bright enough to obtain a spectrum for spectral typing, TiDES will
attempt do so. Otherwise TiDES will target the host galaxy of the
transient to obtain a spectroscopic redshift for type Ia SN cosmology.

The second difference between OzDES and TiDES is that TiDES will have
two components, a wide field component and a deep field component. The
wide field follow-up will cover the majority of the southern sky,
where low redshift transients and host galaxies will be targeted with
a maximum exposure time of two hours. The second component to TiDES
will be spectroscopic follow-up in the deep fields of 4MOST, where
higher redshift live transients will be observed, the AGN
reverberation mapping will take place, and where one can repeatedly
target SN host galaxies to build up redshifts on the faintest of
targets.

The 4MOST Deep fields currently correspond to four pointings, each
covering 4.2\, sq. degrees and contained within the LSST deep drilling
fields \citep{Guiglion2019}. LSST data will provide the high precision
photometry and high cadence light curves required for sub-percent
precision cosmology. As three of the DES deep fields (ECDFS, XMM-LSS,
ELAIS S1) are also covered by the 4MOST and LSST DDFs, this offers the
exciting possibility of continued monitoring a subsample of the AGN
observed by OzDES over a 15 year baseline, if the time gap between
OzDES and TiDES can be covered. The higher cadence and longer
observing season will also mean that TiDES will be more sensitive to
AGN with shorter lags and less sensitive to seasonal gaps. Therefore
TiDES will obtain a higher fraction of lags compared to OzDES.

With more than 4 times the number of fibres, and a field of view that
is only slightly larger that 2dF, TiDES on 4MOST will be able to
observe SN hosts in the deep fields longer without running out of
fibres. Given the higher throughput, broader spectral coverage, and
higher spectral resolution of 4MOST compared to AAOmega, and the
better observing conditions on Cerro Paranal compared to Siding
Spring, TiDES will be able to obtain host galaxy redshifts more
quickly and to a fainter magnitude limit than OzDES.

\section{Summary}

Over six consecutive observing seasons, starting in 2013, OzDES
obtained over 375,000 spectra of almost 39,000 sources in the ten DES
deep fields. These spectra were obtained with the 2dF fibre positioner
and AAOmega spectrograph on the AAT. Over that time, each DES SN field
was observed between 18 and 25 times.

In these fields, OzDES has obtained redshifts for almost 7,000
galaxies that hosted a transient, and classified almost 270
transients. The strategy of targeting each field repeatedly has
enabled redshifts to be obtained for sources as faint as
$r_{\mathrm{AB}}\sim 24$. We show that one is not limited by a noise
floor, even after two days of exposure, and that one can reach high
spectroscopic completeness at these faint limits.

The strategy of targeting each field repeatedly has allowed OzDES to
monitor 771 AGN. These data will be used to measure the lag between
the continuum from the accretion disk and the broad lines from the
broad line emission region. We anticipate that we will recover lags
and derive black hole masses for approximately 30\% of the AGN we
monitored.

The OzDES observing strategy allowed a number of science programmes to
run in parallel. Just over 60\% of the 192,000 fibre hours were
allocated to monitoring AGN and obtaining redshifts for galaxies that
hosted a transient. The remaining 40\% was used to obtain redshifts
for wide variety of targets, such as radio galaxies, cluster galaxies,
and galaxies to train photometric redshift algorithms.

OzDES can be used as a template for future surveys that use the next
generation of wide-field multi-object fibre-fed spectroscopic
facilities to target and monitor sources in the LSST deep drilling
fields. These new facilities will have up to an order of magnitude
more fibres than 2dF, and they will enable one to follow a larger
number of fainter sources more frequently and for longer.

\section*{Acknowledgments}

Parts of this research were supported by the Australian Research
Council under grants DP160100930, FL180100168, and FT140101270.

Based in part on data acquired at the Anglo-Australian Telescope,
under programme A/2013B/012. We acknowledge the traditional owners of
the land on which the AAT stands, the Gamilaraay people, and pay our
respects to elders past, present, and emerging.

We acknowledge the support of James Tocknell and Simon O'Toole from
Data Central in helping us prepare OzDES-DR2.

L.G. was funded by the European Union's Horizon 2020 research and
innovation programme under the Marie Sk\l{}odowska-Curie grant agreement
No. 839090.

F.H.P. was supported by an Australian Government Research Training
Programme Scholarship.

Funding for the DES Projects has been provided by the U.S. Department
of Energy, the U.S. National Science Foundation, the Ministry of
Science and Education of Spain, the Science and Technology Facilities
Council of the United Kingdom, the Higher Education Funding Council
for England, the National Center for Supercomputing Applications at
the University of Illinois at Urbana-Champaign, the Kavli Institute of
Cosmological Physics at the University of Chicago, the Center for
Cosmology and Astro-Particle Physics at the Ohio State University, the
Mitchell Institute for Fundamental Physics and Astronomy at Texas A\&M
University, Financiadora de Estudos e Projetos, Funda{\c c}{\~a}o
Carlos Chagas Filho de Amparo {\`a} Pesquisa do Estado do Rio de
Janeiro, Conselho Nacional de Desenvolvimento Cient{\'i}fico e
Tecnol{\'o}gico and the Minist{\'e}rio da Ci{\^e}ncia, Tecnologia e
Inova{\c c}{\~a}o, the Deutsche Forschungsgemeinschaft and the
Collaborating Institutions in the Dark Energy Survey.

The Collaborating Institutions are Argonne National Laboratory, the
University of California at Santa Cruz, the University of Cambridge,
Centro de Investigaciones Energ{\'e}ticas, Medioambientales y
Tecnol{\'o}gicas-Madrid, the University of Chicago, University College
London, the DES-Brazil Consortium, the University of Edinburgh, the
Eidgen{\"o}ssische Technische Hochschule (ETH) Z{\"u}rich, Fermi
National Accelerator Laboratory, the University of Illinois at
Urbana-Champaign, the Institut de Ci{\`e}ncies de l'Espai (IEEC/CSIC),
the Institut de F{\'i}sica d'Altes Energies, Lawrence Berkeley
National Laboratory, the Ludwig-Maximilians Universit{\"a}t
M{\"u}nchen and the associated Excellence Cluster Universe, the
University of Michigan, the National Optical Astronomy Observatory,
the University of Nottingham, The Ohio State University, the
University of Pennsylvania, the University of Portsmouth, SLAC
National Accelerator Laboratory, Stanford University, the University
of Sussex, Texas A\&M University, and the OzDES Membership Consortium.

Based in part on observations at Cerro Tololo Inter-American
Observatory, National Optical Astronomy Observatory, which is operated
by the Association of Universities for Research in Astronomy (AURA)
under a cooperative agreement with the National Science Foundation.

The DES data management system is supported by the National Science
Foundation under Grant Numbers AST-1138766 and AST-1536171.  The DES
participants from Spanish institutions are partially supported by
MINECO under grants AYA2015-71825, ESP2015-66861, FPA2015-68048,
SEV-2016-0588, SEV-2016-0597, and MDM-2015-0509, some of which include
ERDF funds from the European Union. IFAE is partially funded by the
CERCA programme of the Generalitat de Catalunya.  Research leading to
these results has received funding from the European Research Council
under the European Union's Seventh Framework Program (FP7/2007-2013)
including ERC grant agreements 240672, 291329, and 306478.  We
acknowledge support from the Brazilian Instituto Nacional de Ci\^encia
e Tecnologia (INCT) e-Universe (CNPq grant 465376/2014-2).

This manuscript has been authored by Fermi Research Alliance, LLC
under Contract No. DE-AC02-07CH11359 with the U.S. Department of
Energy, Office of Science, Office of High Energy Physics. The United
States Government retains and the publisher, by accepting the article
for publication, acknowledges that the United States Government
retains a non-exclusive, paid-up, irrevocable, world-wide license to
publish or reproduce the published form of this manuscript, or allow
others to do so, for United States Government purposes.

We are grateful for the extraordinary contributions of our CTIO
colleagues and the DECam Construction, Commissioning and Science
Verification teams in achieving the excellent instrument and telescope
conditions that have made this work possible.  The success of this
project also relies critically on the expertise and dedication of the
DES Data Management group.





\bibliographystyle{mnras}
\bibliography{ozdes}{}


\appendix

\section{Author Affiliations}
\input{files/DES-2019-0517_affiliations_list_v20200229.dat}\label{sec:affil}

\section{Observing logs for Y4, Y5 and Y6}

The observing logs for Y4, Y5, and Y6 are shown in Tables
\ref{tab:obsY4}, \ref{tab:obsY5}, and \ref{tab:obsY6},
respectively. They follow the format presented in Yu15 and Ch17. The
MaxVis field \citep{Yu2020} was observed for 155 minutes during run 28
on 25 and 26 December 2017.

\begin{table*}
 \caption{OzDES fourth year (Y4) observing log for DES deep fields (c.f. Table 2 of Yu15 and Tables A1, A2, and A3 in Ch17).}
 \label{tab:obsY4}
 \begin{tabular}{lcccccccccccl}
  \hline
   UT Date & Observing run$^a$ & \multicolumn{10}{c}{Total exposure time (minutes)} & Notes for entire run\\
    &  & E1 & E2 & S1 & S2 & C1 & C2 & C3 (deep) & X1 & X2 & X3 (deep)& \\
    \input{files/Y4.dat}
  \hline
 
\multicolumn{13}{l}{\footnotesize$^a$ Run numbering includes runs from 2dFLenS and DEVILS projects, so is not necessarily contiguous.}\\
\multicolumn{13}{l}{\footnotesize$^b$ Bright object backup programme for poor conditions, not counted toward final total.}\\

 \end{tabular}
\end{table*}

\begin{table*}
 \caption{OzDES fifth year (Y5) observing log for DES deep fields (c.f. with Table \protect{\ref{tab:obsY4}}).}
 \label{tab:obsY5}
 \begin{tabular}{lcccccccccccl}
  \hline
   UT Date & Observing run$^a$ & \multicolumn{10}{c}{Total exposure time (minutes)} & Notes for entire run\\
    &  & E1 & E2 & S1 & S2 & C1 & C2 & C3 (deep) & X1 & X2 & X3 (deep)& \\
  \input{files/Y5.dat}
   \hline
\multicolumn{13}{l}{\footnotesize$^a$ Run numbering includes runs from 2dFLenS and DEVILS projects, so is not necessarily contiguous.}\\
\multicolumn{13}{l}{\footnotesize$^b$ Bright object backup programme for poor conditions, not counted toward final total.}\\

  \end{tabular}
\end{table*}

\begin{table*}
 \caption{OzDES sixth year (Y6) observing log for DES deep fields (c.f. with Table \protect{\ref{tab:obsY4}}).}
 \label{tab:obsY6}
 \begin{tabular}{lcccccccccccl}
  \hline
   UT Date & Observing run$^a$ & \multicolumn{10}{c}{Total exposure time (minutes)} & Notes for entire run\\
    &  & E1 & E2 & S1 & S2 & C1 & C2 & C3 (deep) & X1 & X2 & X3 (deep)& \\
  \input{files/Y6.dat}
   \hline
\multicolumn{13}{l}{\footnotesize$^a$ Run numbering includes runs from 2dFLenS and DEVILS projects, so is not necessarily contiguous.}\\
\multicolumn{13}{l}{\footnotesize$^b$ Bright object backup programme for poor conditions, not counted toward final total.}\\

 \end{tabular}

\end{table*}

\section{The second OzDES Data release - OzDES-DR2}\label{sec:DR}


The second OzDES data release, OzDES-DR2, consists of the OzDES
redshift catalogue and the reduced spectra.  The data are available
from Data Central\footnote{\url{https://datacentral.org.au}}.

\subsection{Redshift Catalogue}

The OzDES-DR2 redshift catalogue is a FITS binary table containing
almost 38,700 entries, of which almost 30,000 have a redshift with
quality flag 3 or greater. The columns in the catalogue are described
in Table~\ref{tab:OZDES-GRC}.

The transient type is sourced from the 34 ATels
\citep[e.g.,][]{Calcino2018} that OzDES published over 6
years. \cite{DAndrea2018} provides a description off how transients
are classified.

When available, we add comments made by redshifters. Not all sources
have comments, as adding a comment was left to the discretion of the
redshifter.

Many sources were assigned multiple types. For example, sources that
were assigned the LRG type were often assigned as redMaGiC sources and
visa-versa. Sources observed during the OzDES pre-survey (runs 001 and
002) were not given an object type.

\begin{table}
 \caption{Description of the columns in the OzDES-DR2 redshift catalogue.}
 \label{tab:OZDES-GRC}
 \begin{tabular}{lll}
  \hline
   Name & Description & Units \\
   \hline
   OzDES ID & Official DES name or an assigned OzDES name & ...  \\
      & if an official DES name is unavailable       & ...  \\
   Alpha J2000 & Right Ascension                             & deg  \\
   Delta J2000 & Declination                                & deg  \\
   rmag & $r_\mathrm{AB}$ band magnitude                          & ... \\
   $z$ &  Redshift; -9.99 means no redshift                                    & ... \\
   qop (Q) & Redshift quality flag                  & ... \\
   Object types & Assigned types                    & ... \\
   Transient type & Type of transient (e.g. SN Ia, SN II, etc.); Assigned to {\it None} if no type & ...\\
   Comment & Comments from redshifters & ...\\
   \hline
 \end{tabular}
\end{table}

\subsection{Spectra}

All spectra are provided as multi-extension FITS. For targets observed
up to and including run 40, there is usually one FITS file per
target. If a target was also observed during run 41, which used a
redder setting for AAOmega, we provide a second FITS file containing
the data from that run combined with data from earlier runs. A small
number of targets in run 41 were only observed in that run. In this
case, there is just one FITS file per target.

Within each FITS file, there are extensions containing the stacked
spectrum and the spectra that went into the stack. Also included are
the variances and bad pixel masks for each spectrum. The extensions
are described in Table~\ref{tab:extensions}. The individual spectra
that go into the stack are ordered by date, from the one taken first
to the last.

All spectra have been shifted to the heliocentric reference frame. 

\begin{table}
 \caption{Description of FITS extensions in OzDES-DR2.}
 \label{tab:extensions}
 \begin{tabular}{ll}
  \hline
   Extension & Description\\
    \hline
   Primary   & Stacked spectrum \\
   1         & Variance spectrum\\
   2         & Mask \\
   3         & Spectrum of first epoch \\
   4         & Variance spectrum of first epoch \\
   5         & Mask for first epoch\\
   $\vdots$  & $\vdots$ \\
   $3i$      & Spectrum of {\it i}th epoch \\
   $3i+1$    & Variance spectrum of {\it i}th epoch \\
   $3i+2$    & Mask for {\it i}th epoch\\  
  \hline
   \hline
 \end{tabular}
\end{table}

The keywords in the FITS headers are described in Table~\ref{tab:keywords}. 

\begin{table*}
 \caption{Description of FITS headers in OzDES-DR2.}
 \label{tab:keywords}
 \begin{tabular}{ll}
  \hline\hline
  \multicolumn{2}{l}{Primary Header$^a$}\\
  \hline
  DO\_HELIO & An internal flag recognised by the redshift code MARZ that ensures that no helio-centric \\
            & correction is applied by MARZ when set to false\\
  NDATE & The date the data were combined \\
  SOURCE & The original source name \\
  VERCOADD & The version of the coadd pipeline that stacks the reduced data\\
  RMAG & $r_{\mathrm AB}$ band magnitude in a 2\arcsec\ aperture\\
  TYPEn & Target type \\
  Z & redshift; -9.99 means no redshift \\
  QOP & redshift quality flag \\
  \hline
   \multicolumn{2}{l}{Individual epochs$^b$}\\
   \hline
   REDZP$^d$ & The zero point in the $r$-band. Measured from the spectra of F stars that were observed \\
         & at the same time as the source \\
   
   REDSENSF & The sensitivity function used in the red arm\\
   BLUEZP & The zero point in the $g$-band. Measured from the spectra of F stars that were observed \\
         & at the same time as the source\\
   BLUSENSF & The sensitivity function used in the blue arm\\
   QC & Quality control flag\\
   VERSION & The version of the pipeline that processes the raw data and produces single epoch reduced spectra \\
   INCOADD & True if the epoch is included in the coadd \\
   PIVOT & The 2dF fibre number\\
   XPLATE & The X position on the plate (microns)\\
   YPLATE & The Y position on the plate (microns)\\
   ORIGTARG & The name of the target used in this observation\\
   SOURCEF &  The raw data filename (red arm only)\\
   FLXSCALE & The scale factor that was applied in the co-adding\\
   FLSCTYPE & The method used to compute scale factor\\
   RMAG & $r_{\mathrm AB}$ in the input target catalogue\\
   \hline
   \multicolumn{2}{l}{Individual epochs$^c$}\\
   \hline
   EXPOSED & Exposure time \\
   DICHROIC & Dichroic used - x6700 for run041, x5700 otherwise \\
   SOURCE & 2dF field plate - either plate 0 or plate 1\\
   UTMJD & Modified  Julian Date \\
   ZDSTART & Zenith Distance at the start of the exposure \\
   ZDEND & Zenith Distance at the end of the exposure \\
   
   \hline\hline
   \multicolumn{2}{l}{$^a$ Keywords that are in the FITS standard are not listed.}\\
   \multicolumn{2}{l}{$^b$ Keywords that are inserted when the data are co-added.}\\
   \multicolumn{2}{l}{$^c$ Keywords that are part of the header produced at the telescope and keywords. Not all keywords are listed.}\\
   \multicolumn{2}{l}{$^d$ Zero points are defined as $2.5 \log_{10}(f)+m $, where $m$ is the catalog magnitude of the F star and $f$ is the observed}\\ 
   \multicolumn{2}{l}{spectrum integrated over the $g$ band for the blue arm and the $r$ band for the red arm.}
 \end{tabular}
\end{table*}

\subsection{Data Quality}\label{sec:dataquality}

The quality of the data is characterised in a couple of ways. The
keywords REDZP and BLUEZP give an indication of the observing
conditions at the time the data were taken. They are determined from F
stars that were observed at the same time as the target. For the red
arm, a zero point that is smaller than 31 usually means that
conditions were poor, which could be due to seeing or the presence of
cloud. It could also be due to an exposure that was cut short.

Most of the 375,000 spectra in the data release have been visually
inspected and problematic cases noted in the QC keyword. As this is
not 100\% foolproof, there might be a number of problematic cases that
have been missed. A value for QC that differs from ``ok'' indicates
that there is a problem with the spectrum. The nature of the problem
is noted in the value of the QC keyword and is described in
Table~\ref{tab:artefacts}.

Some of the problems are caused by artifacts that affect a limited
wavelength range. Others affect the whole spectrum. The region of the
spectrum affected by the artifact is noted in the masks. A value of 1
in the mask marks the region of poor quality data.

Examples of what the artifacts look like are available from the
\href{https://docs.datacentral.org.au/ozdes/overview/dr2/}{OzDES-DR2
  documentation page} at Data Central.

\begin{table*}
 \caption{Description of values contained in the QC keyword.}
 \label{tab:artefacts}
 \begin{tabular}{ll}
  \hline
  QC value & Description \\
  \hline
  700nmLED & Contamination from an LED in the gripper arm. This was discovered a few years, \\
           &  into the survey, so it mostly affects early data. \\
  redDip   & Most likely associated to charge traps in the old AAOmega red CCD. Affects data taken prior \\
           & to run007. \\
  doubleDip\_1 & Most likely associated to charge traps in the old AAOmega red CCD. Affects data taken prior \\
           & to run007. \\
  doubleDip\_2 & Most likely associated to charge traps in the old AAOmega red CCD. Affects data taken prior \\
           & to run007. \\
  doubleHump & Origin unknown. \\
  fib388p0 & Issues with fibre 388 on plate 0. Cause unknown. \\
  fib94p1 & Issues with fibre 94 on plate 1. Cause unknown. \\
  redfail & Failed reduction in red arm. \\
  redPCAFail & Failed PCA sky subtraction in the red arm. \\
  bluefail & Failed reduction in blue arm. \\
  failedReduction & Failed reduction, both arms. \\
  redPCAfail & Failed PCA sky subtraction in red arm. \\
  brightStarBlue & A bright star in a nearby fibre causes the reduction in the blue arm to fail or contaminates the\\ & extracted spectrum in the blue arm.\\
  brightStarRed & As above, but for the red arm.\\
  brightStarAll & As above, but for the entire spectrum.\\
  fringed & Fringes detected. \\
  poorConditions & Poor observing conditions. \\
  blueSkySubFail & Sky subtraction failed in blue arm\\
  blueFeature\_1 & A blue feature of unkown origin\\
  IRcontamination & Contamination in red arm of unknown origin\\
  failedReduction & Processing failed for unspecified reason \\
  
  \hline
 \end{tabular}
\end{table*}

\section{Source types for the MaxVis field}\label{sec:MaxVis}

OzDES targeted the DES MaxVis field towards the end of Y5. Centred at
RA=$97.5\degree$ and DEC=$-58.75\degree$, it is one of the DES
standard star fields and is observable during the entire DES observing
season.  Listed in Table~\ref{tab:MaxVis} is a description of the
target types together with the number observed.

In more detail, stellar-like objects were selected on the basis of
their location in $r-i$ vs $g-r$ colour-colour diagram or on the
amount of variability that they showed. Targets with more variability
were given higher priority. Some of the variable sources are AGN.

The three tail sources were located in a small tail-like feature in
the colour-colour diagram near to where the stellar sequence turns up
with redder stars. All three sources were galaxies.

QSO candidates were selected on the basis of their colour, with QSOuv
representing sources that were red in $r-i$, but blue in $g-r$, QSOHz
representing QSOs that were likely to be at high redshift, and QSOc
representing sources that were selected by the Cambridge QSO group.

At lower priority were blue stars (BBstar), of which three were white
dwarfs and two were AGN. Lowest priority were other randomly selected
stars (RNDstars).

\begin{table*}
 \caption{Description of target types in the DES MaxVis field.}
 \label{tab:MaxVis}
 \begin{tabular}{lrl}
  \hline
  Type & Number of sources & Description \\
  \hline
  \multicolumn{3}{l}{High Priority targets}  \\ 
  \hline
   QSOHz & 49 & High-$z$ quasar candidates\\
   QSOuv & 14 & uv-excess quasar candidates\\
   QSOc  & 67 & quasar candidates selected from the Cambridge QSO group\\
   VarObj1 & 94 & Variable Object candidate (high priority)\\
   Tail1, Tail2, Tail3 & &  3 star-like objects with colour in the tail-like region\\
  \hline
  \multicolumn{3}{l}{Low Priority targets}  \\ 
  \hline
   BBstar & 72 &  stars selected on the basis of blue colours in g-r and r-i\\
   VarObj2 & 55 & Variable Object candidate (lower priority)\\
    \hline
   \multicolumn{3}{l}{Lowest Priority targets}  \\ 
   \hline
   RNDstars &1 & Random star-like objects\\
  \hline
 \end{tabular}
\end{table*}


\bsp	
\label{lastpage}

\end{document}

%% file: files/Y4.dat
&&&&&&&&&&&&\\
2016-08-25	&024	& 80	&-	&-	&-	&-	&-	&-	&-	&-	&-	& 0.5/1.5 clear nights\\
2016-08-26	& 	&-	& 90	&-	&-	&-	&-	&-	&-	&-	&-	&\\
2016-08-27	& 	&-	&-	&-	&-	&-	&-	&-	&-	&-	&110	&\\
2016-08-28	& 	&-	&-	&-	& 85	&-	&-	&-	&-	&-	&-	&\\
2016-08-29	& 	&180	&-	&-	&-	&-	&-	& 80	& 80	&-	&-	&\\
&&&&&&&&&&&&\\
2016-09-25	&025	&-	& 80	&-	&-	&-	& 20	&120	&-	&-	&120	& 3.5/4.5 clear nights\\
2016-09-26	& 	& 80	&-	&-	&-	& 30	&120	&-	&120	&-	&-	&\\
2016-09-27	& 	&-	& 80	& 40	&-	& 80	&-	&-	&-	&120	&-	&\\
2016-10-05	& 	&-	& 40	&100	&120	&-	&-	&-	&-	&-	&-	&\\
&&&&&&&&&&&&\\
2016-10-31	&026	&-	&-	&-	&-	&-	&-	& 80	&-	&-	&120	& 3.3/3.5 clear nights\\
2016-11-01	& 	&120	&-	&120	&-	&-	&-	& 60	&-	&-	&120	&\\
2016-11-02	& 	&-	&120	&-	&120	&-	&-	&-	& 80	&-	&-	&\\
2016-11-03	& 	& 80	&-	&-	&-	&120	&110	&-	&-	&120	&-	&\\
&&&&&&&&&&&&\\
2016-11-25	&027	&120	&-	&-	&-	& 40	&-	&120	&-	&-	&120	& 5.0/5.0 clear nights\\
2016-11-26	& 	&-	&120	&120	&-	&-	&120	&-	&-	&-	&-	&\\
2016-11-27	& 	&-	&-	&-	&120	&120	&-	&-	&120	&-	&-	&\\
2016-11-28	& 	&-	&-	&120	&-	&-	& 68	& 80	&-	&120	&-	&\\
2016-11-29	& 	&120	&-	&-	&-	&-	&153	&-	&-	&-	&120	&\\
&&&&&&&&&&&&\\
2016-12-25	&028	&-	&-	&-	&-	&-	& 80	& 80	&-	&-	&120	& 2.8/4.0 clear nights\\
2016-12-26	& 	&-	&-	& 80	&-	& 80	&-	&-	&-	& 80	&-	&\\
2016-12-27	& 	&-	&-	&-	& 80	&-	&-	& 80	& 80	&-	&-	&\\
&&&&&&&&&&&&\\
All Y4	&Total(min) &780	&530	&580	&525	&470	&671	&700	&480	&440	&830	&\\

%% file: files/Y5.dat
&&&&&&&&&&&&\\
2017-08-21	&030	& 80	& 80	&-	&-	&-	&-	& 80	&-	&-	& 80	& 4.0/4.2 clear nights\\
2017-08-22	& 	& 80	& 80	& 80	&-	&-	&120	&-	&-	&-	&-	&\\
2017-08-23	& 	& 40	& 80	&-	&-	& 80	&-	&-	&-	& 80	&-	&\\
2017-08-24	& 	& 80	& 80	&-	&120	&-	&-	&-	& 80	&-	&-	&\\
2017-08-25	& 	& 80	& 80	&-	&-	&-	&-	&120	&-	&-	& 80	&\\
2017-08-26	& 	& 34	& 40	&-	&-	& 49	&-	&-	& 25	&-	&-	&\\
2017-08-27	& 	&-	&-	&-	& 80	&-	&-	&-	& 80	& 80	&-	&\\
&&&&&&&&&&&&\\
2017-10-17	&032	&-	&-	&-	&-	& 40	&-	& 75	&-	&-	&-	& 4.3/6.0 clear nights\\
2017-10-18	& 	&120	&-	&120	&-	&-	&103	&-	&-	&-	&120	&\\
2017-10-21	& 	&-	&120	&-	&130	& 72	&-	&-	&-	&120	&-	&\\
2017-10-22	& 	& 93	& 80	& 80	&-	&-	&-	& 80	&120	&-	&-	&\\
2017-10-23	& 	&-	&105	&-	&120	&-	&-	&102	&-	&-	&120	&\\
&&&&&&&&&&&&\\
2017-11-16	&033	& 80	&-	&-	&-	& 60	&-	& 80	&-	&-	&120	& 2.8/4.5 clear nights\\
2017-11-19	& 	&-	& 80	&-	&-	& 30	&120	&-	& 80	& 80	&-	&\\
2017-11-20	& 	&-	&-	&120	&120	& 40	&-	& 80	&-	&-	&-	&\\
&&&&&&&&&&&&\\
2017-12-18	&034	&-	&-	&-	&-	&-	&-	&100	&-	&-	& 80	& 3.3/4.0 clear nights\\
2017-12-19	& 	&-	&-	&-	&-	& 80	&-	&-	&-	& 80	&-	&\\
2017-12-21	& 	&-	&-	&-	&-	&-	&107	&-	&-	&-	& 50	&\\
2017-12-22	& 	& 80	&-	&-	&-	&-	&110	&-	&-	&-	&-	&\\
2017-12-23	& 	&-	& 80	&-	&-	&-	&-	&-	&100	&-	&-	&\\
2017-12-24	& 	&-	&-	& 70	&-	& 80	&-	&-	&-	&-	&-	&\\
2017-12-25	& 	&-	&-	&-	& 87	&-	& 80	&-	&-	&-	&-	&\\
&&&&&&&&&&&&\\
All Y5	&Total(min) &768	&905	&470	&657	&531	&640	&717	&485	&440	&650	&\\

%% file: files/Y6.dat
&&&&&&&&&&&&\\
2018-06-15	&036	&-	& 80	&-	&-	&-	&-	&-	&-	&-	&-	& 0.6/1.5 clear nights\\
2018-06-24	& 	&150	& 80	&-	&-	&-	&-	&-	&-	&-	&-	&\\
&&&&&&&&&&&&\\
2018-09-04	&037	& 60	&120	&-	&-	&-	& 40	&120	&-	&-	& 80	& 1.0/3.5 clear nights\\
2018-09-05	& 	&117	&-	&-	&-	&-	&-	&-	&-	&-	&-	&\\
2018-09-06	& 	&-	&-	&-	&-	&-	&-	&-	&-	& 40	&-	&\\
&&&&&&&&&&&&\\
2018-10-06	&038	&120	&-	&120	&-	&-	&-	& 80	&-	&-	&120	& 3.0/3.0 clear nights\\
2018-10-07	& 	&-	&120	&-	&120	&-	&120	&-	&-	&120	&-	&\\
2018-10-08	& 	&120	&-	&-	&-	&120	&-	&120	&120	&-	&-	&\\
&&&&&&&&&&&&\\
2018-11-08	&039	&-	&120	& 40	&-	&-	&-	&120	&-	&-	&120	& 2.0/3.0 clear nights\\
2018-11-09	& 	&120	&-	& 40	&-	&-	&120	&-	&-	&120	&-	&\\
&&&&&&&&&&&&\\
2018-12-01	&040	& 80	&-	& 80	&-	&-	&-	& 80	&-	&-	& 80	& 2.2/3.0 clear nights\\
2018-12-02	& 	&-	& 80	&-	& 80	&-	&120	&-	&-	& 80	&-	&\\
2018-12-03	& 	& 74	&-	&-	&-	& 22	&-	&-	& 71	&-	&-	&\\
&&&&&&&&&&&&\\
2019-01-04	&041	&-	&-	&-	&-	&-	&-	&120	&-	&-	& 80	& 1.7/3.0 clear nights\\
2019-01-06	& 	&-	&-	&-	&-	&-	&-	&120	&-	&-	& 80	&\\
2019-01-08	& 	&-	&-	&-	&-	&-	&-	&120	&-	&-	& 80	&\\
2019-01-09	& 	&-	&-	&-	&-	&-	&-	&-	&-	&-	& 40	&\\
&&&&&&&&&&&&\\
All Y6	&Total(min) &841	&600	&280	&200	&142	&400	&880	&191	&360	&680	&\\

%% file: ozdes.bbl
\begin{thebibliography}{}
\makeatletter
\relax
\def\mn@urlcharsother{\let\do\@makeother \do\$\do\&\do\#\do\^\do\_\do\%\do\~}
\def\mn@doi{\begingroup\mn@urlcharsother \@ifnextchar [ {\mn@doi@}
  {\mn@doi@[]}}
\def\mn@doi@[#1]#2{\def\@tempa{#1}\ifx\@tempa\@empty \href
  {http://dx.doi.org/#2} {doi:#2}\else \href {http://dx.doi.org/#2} {#1}\fi
  \endgroup}
\def\mn@eprint#1#2{\mn@eprint@#1:#2::\@nil}
\def\mn@eprint@arXiv#1{\href {http://arxiv.org/abs/#1} {{\tt arXiv:#1}}}
\def\mn@eprint@dblp#1{\href {http://dblp.uni-trier.de/rec/bibtex/#1.xml}
  {dblp:#1}}
\def\mn@eprint@#1:#2:#3:#4\@nil{\def\@tempa {#1}\def\@tempb {#2}\def\@tempc
  {#3}\ifx \@tempc \@empty \let \@tempc \@tempb \let \@tempb \@tempa \fi \ifx
  \@tempb \@empty \def\@tempb {arXiv}\fi \@ifundefined
  {mn@eprint@\@tempb}{\@tempb:\@tempc}{\expandafter \expandafter \csname
  mn@eprint@\@tempb\endcsname \expandafter{\@tempc}}}

\bibitem[\protect\citeauthoryear{{Angus} et~al.,}{{Angus}
  et~al.}{2019}]{Angus2019}
{Angus} C.~R.,  et~al., 2019, \mn@doi [\mnras] {10.1093/mnras/stz1321}, \href
  {https://ui.adsabs.harvard.edu/abs/2019MNRAS.487.2215A} {487, 2215}

\bibitem[\protect\citeauthoryear{{Astier} et~al.,}{{Astier}
  et~al.}{2006}]{Astier2006}
{Astier} P.,  et~al., 2006, \mn@doi [\aap] {10.1051/0004-6361:20054185}, \href
  {http://adsabs.harvard.edu/abs/2006A%26A...447...31A} {447, 31}

\bibitem[\protect\citeauthoryear{{Bernstein} et~al.,}{{Bernstein}
  et~al.}{2012}]{Bernstein2012}
{Bernstein} J.~P.,  et~al., 2012, \mn@doi [\apj] {10.1088/0004-637X/753/2/152},
  \href {http://adsabs.harvard.edu/abs/2012ApJ...753..152B} {753, 152}

\bibitem[\protect\citeauthoryear{{Betoule} et~al.,}{{Betoule}
  et~al.}{2014}]{Betoule2014}
{Betoule} M.,  et~al., 2014, \mn@doi [\aap] {10.1051/0004-6361/201423413},
  \href {http://adsabs.harvard.edu/abs/2014A%26A...568A..22B} {568, A22}

\bibitem[\protect\citeauthoryear{{Bonnett} et~al.,}{{Bonnett}
  et~al.}{2016}]{Bonnett2016}
{Bonnett} C.,  et~al., 2016, \mn@doi [\prd] {10.1103/PhysRevD.94.042005}, \href
  {http://adsabs.harvard.edu/abs/2016PhRvD..94d2005B} {94, 042005}

\bibitem[\protect\citeauthoryear{{Calcino} \& {Davis}}{{Calcino} \&
  {Davis}}{2017}]{Calcino2017}
{Calcino} J.,  {Davis} T.,  2017, \mn@doi [\jcap]
  {10.1088/1475-7516/2017/01/038}, \href
  {https://ui.adsabs.harvard.edu/abs/2017JCAP...01..038C} {2017, 038}

\bibitem[\protect\citeauthoryear{{Calcino} et~al.,}{{Calcino}
  et~al.}{2018}]{Calcino2018}
{Calcino} J.,  et~al., 2018, The Astronomer's Telegram, \href
  {https://ui.adsabs.harvard.edu/abs/2018ATel11146....1C} {11146, 1}

\bibitem[\protect\citeauthoryear{{Childress} et~al.,}{{Childress}
  et~al.}{2017}]{Childress2017}
{Childress} M.~J.,  et~al., 2017, \mn@doi [\mnras] {10.1093/mnras/stx1872},
  \href {http://adsabs.harvard.edu/abs/2017MNRAS.472..273C} {472, 273}

\bibitem[\protect\citeauthoryear{{Croom}, {Saunders}  \& {Heald}}{{Croom}
  et~al.}{2004}]{Croom2004}
{Croom} S.,  {Saunders} W.,   {Heald} R.,  2004, Anglo-Australian Observatory
  Epping Newsletter, \href {http://adsabs.harvard.edu/abs/2004AAONw.106...12C}
  {106, 12}

\bibitem[\protect\citeauthoryear{{D'Andrea} et~al.,}{{D'Andrea}
  et~al.}{2018}]{DAndrea2018}
{D'Andrea} C.~B.,  et~al., 2018, arXiv e-prints, \href
  {https://ui.adsabs.harvard.edu/abs/2018arXiv181109565D} {}

\bibitem[\protect\citeauthoryear{{DESI Collaboration}}{{DESI
  Collaboration}}{2016}]{DESI2016}
{DESI Collaboration} 2016, arXiv e-prints, \href
  {http://adsabs.harvard.edu/abs/2016arXiv161100036D} {}

\bibitem[\protect\citeauthoryear{{Dark Energy Survey}}{{Dark Energy
  Survey}}{2016}]{DES2016}
{Dark Energy Survey} 2016, \mn@doi [\mnras] {10.1093/mnras/stw641}, \href
  {http://adsabs.harvard.edu/abs/2016MNRAS.460.1270D} {460, 1270}

\bibitem[\protect\citeauthoryear{{Dark Energy Survey}}{{Dark Energy
  Survey}}{2019a}]{DES2019b}
{Dark Energy Survey} 2019a, \mn@doi [\prl] {10.1103/PhysRevLett.122.171301},
  \href {https://ui.adsabs.harvard.edu/abs/2019PhRvL.122q1301A} {122, 171301}

\bibitem[\protect\citeauthoryear{{Dark Energy Survey}}{{Dark Energy
  Survey}}{2019b}]{DES2019a}
{Dark Energy Survey} 2019b, \mn@doi [\apjl] {10.3847/2041-8213/ab04fa}, \href
  {https://ui.adsabs.harvard.edu/abs/2019ApJ...872L..30A} {872, L30}

\bibitem[\protect\citeauthoryear{{Davies} et~al.,}{{Davies}
  et~al.}{2018}]{Davies2018}
{Davies} L.~J.~M.,  et~al., 2018, \mn@doi [\mnras] {10.1093/mnras/sty1553},
  \href {http://adsabs.harvard.edu/abs/2018MNRAS.480..768D} {480, 768}

\bibitem[\protect\citeauthoryear{{Dawson} et~al.,}{{Dawson}
  et~al.}{2009}]{Dawson2009}
{Dawson} K.~S.,  et~al., 2009, \mn@doi [\aj] {10.1088/0004-6256/138/5/1271},
  \href {http://adsabs.harvard.edu/abs/2009AJ....138.1271D} {138, 1271}

\bibitem[\protect\citeauthoryear{{Diehl} et~al.,}{{Diehl}
  et~al.}{2018}]{Diehl2018}
{Diehl} H.~T.,  et~al., 2018, in \procspie. p. 107040D,
  \mn@doi{10.1117/12.2312113}

\bibitem[\protect\citeauthoryear{{Flaugher} et~al.,}{{Flaugher}
  et~al.}{2015}]{Flaugher2015}
{Flaugher} B.,  et~al., 2015, \mn@doi [\aj] {10.1088/0004-6256/150/5/150},
  \href {https://ui.adsabs.harvard.edu/abs/2015AJ....150..150F} {150, 150}

\bibitem[\protect\citeauthoryear{{Franzen} et~al.,}{{Franzen}
  et~al.}{2015}]{Franzen2015}
{Franzen} T.~M.~O.,  et~al., 2015, \mn@doi [\mnras] {10.1093/mnras/stv1866},
  \href {https://ui.adsabs.harvard.edu/abs/2015MNRAS.453.4020F} {453, 4020}

\bibitem[\protect\citeauthoryear{{Gschwend} et~al.,}{{Gschwend}
  et~al.}{2018}]{Gschwend2018}
{Gschwend} J.,  et~al., 2018, \mn@doi [Astronomy and Computing]
  {10.1016/j.ascom.2018.08.008}, \href
  {http://adsabs.harvard.edu/abs/2018A%26C....25...58G} {25, 58}

\bibitem[\protect\citeauthoryear{{Guiglion} et~al.,}{{Guiglion}
  et~al.}{2019}]{Guiglion2019}
{Guiglion} G.,  et~al., 2019, \mn@doi [The Messenger]
  {10.18727/0722-6691/5119}, \href
  {https://ui.adsabs.harvard.edu/abs/2019Msngr.175...17G} {175, 17}

\bibitem[\protect\citeauthoryear{{Hinton}, {Davis}, {Lidman}, {Glazebrook}  \&
  {Lewis}}{{Hinton} et~al.}{2016}]{Hinton2016}
{Hinton} S.~R.,  {Davis} T.~M.,  {Lidman} C.,  {Glazebrook} K.,   {Lewis}
  G.~F.,  2016, \mn@doi [Astronomy and Computing]
  {10.1016/j.ascom.2016.03.001}, \href
  {http://adsabs.harvard.edu/abs/2016A%26C....15...61H} {15, 61}

\bibitem[\protect\citeauthoryear{{Hoormann} et~al.,}{{Hoormann}
  et~al.}{2019}]{Hoormann2019}
{Hoormann} J.~K.,  et~al., 2019, \mn@doi [\mnras] {10.1093/mnras/stz1539},
  \href {https://ui.adsabs.harvard.edu/abs/2019MNRAS.487.3650H} {487, 3650}

\bibitem[\protect\citeauthoryear{{Jacobs} et~al.,}{{Jacobs}
  et~al.}{2019}]{Jacobs2019}
{Jacobs} C.,  et~al., 2019, \mn@doi [\mnras] {10.1093/mnras/stz272}, \href
  {https://ui.adsabs.harvard.edu/abs/2019MNRAS.484.5330J} {484, 5330}

\bibitem[\protect\citeauthoryear{{Johnson} et~al.,}{{Johnson}
  et~al.}{2017}]{Johnson2017}
{Johnson} A.,  et~al., 2017, \mn@doi [\mnras] {10.1093/mnras/stw3033}, \href
  {http://adsabs.harvard.edu/abs/2017MNRAS.465.4118J} {465, 4118}

\bibitem[\protect\citeauthoryear{{Khain} et~al.,}{{Khain}
  et~al.}{2018}]{Khain2018}
{Khain} T.,  et~al., 2018, \mn@doi [\aj] {10.3847/1538-3881/aaeb2a}, \href
  {https://ui.adsabs.harvard.edu/abs/2018AJ....156..273K} {156, 273}

\bibitem[\protect\citeauthoryear{{King} et~al.,}{{King}
  et~al.}{2015}]{King2015}
{King} A.~L.,  et~al., 2015, \mn@doi [\mnras] {10.1093/mnras/stv1718}, \href
  {http://adsabs.harvard.edu/abs/2015MNRAS.453.1701K} {453, 1701}

\bibitem[\protect\citeauthoryear{{LSST Science Collaboration}}{{LSST Science
  Collaboration}}{2017}]{LSST2017}
{LSST Science Collaboration} 2017, arXiv e-prints, \href
  {http://adsabs.harvard.edu/abs/2017arXiv170804058L} {}

\bibitem[\protect\citeauthoryear{{Lawrence} et~al.,}{{Lawrence}
  et~al.}{2018}]{Lawrence2018}
{Lawrence} J.,  et~al., 2018, in \procspie. p. 10702A6,
  \mn@doi{10.1117/12.2314178}

\bibitem[\protect\citeauthoryear{{Li} et~al.,}{{Li} et~al.}{2019}]{Li2019}
{Li} T.~S.,  et~al., 2019, \mn@doi [\mnras] {10.1093/mnras/stz2731}, \href
  {https://ui.adsabs.harvard.edu/abs/2019MNRAS.490.3508L} {490, 3508}

\bibitem[\protect\citeauthoryear{{Marshall} et~al.,}{{Marshall}
  et~al.}{2019}]{Marshall2019}
{Marshall} J.,  et~al., 2019, \mn@doi [\apj] {10.3847/1538-4357/ab3653}, \href
  {https://ui.adsabs.harvard.edu/abs/2019ApJ...882..177M} {882, 177}

\bibitem[\protect\citeauthoryear{{Mudd} et~al.,}{{Mudd}
  et~al.}{2018}]{Mudd2018}
{Mudd} D.,  et~al., 2018, \mn@doi [\apj] {10.3847/1538-4357/aac9bb}, \href
  {https://ui.adsabs.harvard.edu/abs/2018ApJ...862..123M} {862, 123}

\bibitem[\protect\citeauthoryear{{Nord} et~al.,}{{Nord}
  et~al.}{2016}]{Nord2016}
{Nord} B.,  et~al., 2016, \mn@doi [\apj] {10.3847/0004-637X/827/1/51}, \href
  {https://ui.adsabs.harvard.edu/abs/2016ApJ...827...51N} {827, 51}

\bibitem[\protect\citeauthoryear{{Palmese} et~al.,}{{Palmese}
  et~al.}{2017}]{Palmese2017}
{Palmese} A.,  et~al., 2017, \mn@doi [\apjl] {10.3847/2041-8213/aa9660}, \href
  {http://adsabs.harvard.edu/abs/2017ApJ...849L..34P} {849, L34}

\bibitem[\protect\citeauthoryear{{Pierre} et~al.,}{{Pierre}
  et~al.}{2016}]{Pierre2016}
{Pierre} M.,  et~al., 2016, \mn@doi [\aap] {10.1051/0004-6361/201526766}, \href
  {http://adsabs.harvard.edu/abs/2016A%26A...592A...1P} {592, A1}

\bibitem[\protect\citeauthoryear{{Prat} et~al.,}{{Prat}
  et~al.}{2018}]{Sanchez2018}
{Prat} J.,  et~al., 2018, \mn@doi [\prd] {10.1103/PhysRevD.98.042005}, \href
  {http://adsabs.harvard.edu/abs/2018PhRvD..98d2005P} {98, 042005}

\bibitem[\protect\citeauthoryear{{Pursiainen} et~al.,}{{Pursiainen}
  et~al.}{2018}]{Pursiainen2018}
{Pursiainen} M.,  et~al., 2018, \mn@doi [\mnras] {10.1093/mnras/sty2309}, \href
  {https://ui.adsabs.harvard.edu/abs/2018MNRAS.481..894P} {481, 894}

\bibitem[\protect\citeauthoryear{{Rozo} et~al.,}{{Rozo}
  et~al.}{2016}]{Rozo2016}
{Rozo} E.,  et~al., 2016, \mn@doi [\mnras] {10.1093/mnras/stw1281}, \href
  {http://adsabs.harvard.edu/abs/2016MNRAS.461.1431R} {461, 1431}

\bibitem[\protect\citeauthoryear{{S{\'a}nchez} et~al.,}{{S{\'a}nchez}
  et~al.}{2014}]{Sanchez2014}
{S{\'a}nchez} C.,  et~al., 2014, \mn@doi [\mnras] {10.1093/mnras/stu1836},
  \href {http://adsabs.harvard.edu/abs/2014MNRAS.445.1482S} {445, 1482}

\bibitem[\protect\citeauthoryear{{Smith} et~al.,}{{Smith}
  et~al.}{2018}]{Smith2018}
{Smith} M.,  et~al., 2018, \mn@doi [\apj] {10.3847/1538-4357/aaa126}, \href
  {https://ui.adsabs.harvard.edu/abs/2018ApJ...854...37S} {854, 37}

\bibitem[\protect\citeauthoryear{{Swann} et~al.,}{{Swann}
  et~al.}{2019}]{Swann2019}
{Swann} E.,  et~al., 2019, \mn@doi [The Messenger] {10.18727/0722-6691/5129},
  \href {http://adsabs.harvard.edu/abs/2019Msngr.175...58S} {175, 58}

\bibitem[\protect\citeauthoryear{{Tamura} et~al.,}{{Tamura}
  et~al.}{2018}]{Tamura2018}
{Tamura} N.,  et~al., 2018, in Ground-based and Airborne Instrumentation for
  Astronomy VII. p. 107021C, \mn@doi{10.1117/12.2311871}

\bibitem[\protect\citeauthoryear{{The MSE Science Team} et~al.,}{{The MSE
  Science Team} et~al.}{2019}]{MSE2019}
{The MSE Science Team} et~al., 2019, arXiv e-prints, \href
  {http://adsabs.harvard.edu/abs/2019arXiv190404907T} {}

\bibitem[\protect\citeauthoryear{{Vargas-Magana}, {Brooks}, {Levi}  \&
  {Tarle}}{{Vargas-Magana} et~al.}{2019}]{Vargus-Magana2019}
{Vargas-Magana} M.,  {Brooks} D.~D.,  {Levi} M.~M.,   {Tarle} G.~G.,  2019,
  arXiv e-prints, \href {http://adsabs.harvard.edu/abs/2019arXiv190101581V} {}

\bibitem[\protect\citeauthoryear{{Webb} et~al.,}{{Webb}
  et~al.}{2015}]{Webb2015}
{Webb} T.~M.~A.,  et~al., 2015, \mn@doi [\apj] {10.1088/0004-637X/814/2/96},
  \href {http://adsabs.harvard.edu/abs/2015ApJ...814...96W} {814, 96}

\bibitem[\protect\citeauthoryear{{Wilson} et~al.,}{{Wilson}
  et~al.}{2009}]{Wilson2009}
{Wilson} G.,  et~al., 2009, \mn@doi [\apj] {10.1088/0004-637X/698/2/1943},
  \href {http://adsabs.harvard.edu/abs/2009ApJ...698.1943W} {698, 1943}

\bibitem[\protect\citeauthoryear{{Yasuda} et~al.,}{{Yasuda}
  et~al.}{2019}]{Yasuda2019}
{Yasuda} N.,  et~al., 2019, \mn@doi [\pasj] {10.1093/pasj/psz050}, \href
  {https://ui.adsabs.harvard.edu/abs/2019PASJ...71...74Y} {71, 74}

\bibitem[\protect\citeauthoryear{{Yu} et~al.,}{{Yu} et~al.}{2020}]{Yu2020}
{Yu} Z.,  et~al., 2020, \mn@doi [\apjs] {10.3847/1538-4365/ab5e7a}, \href
  {https://ui.adsabs.harvard.edu/abs/2020ApJS..246...16Y} {246, 16}

\bibitem[\protect\citeauthoryear{{Yuan} et~al.,}{{Yuan}
  et~al.}{2015}]{Yuan2015}
{Yuan} F.,  et~al., 2015, \mn@doi [\mnras] {10.1093/mnras/stv1507}, \href
  {http://adsabs.harvard.edu/abs/2015MNRAS.452.3047Y} {452, 3047}

\bibitem[\protect\citeauthoryear{{de Jong} et~al.,}{{de Jong}
  et~al.}{2019}]{deJong2019}
{de Jong} R.~S.,  et~al., 2019, \mn@doi [The Messenger]
  {10.18727/0722-6691/5117}, \href
  {http://adsabs.harvard.edu/abs/2019Msngr.175....3D} {175, 3}

\makeatother
\end{thebibliography}
